\pdfoutput=1

\documentclass[review,10pt]{article}

\usepackage{authblk}                
\usepackage[square,numbers]{natbib} 

\usepackage{hyperref}

\usepackage{graphicx}%
\usepackage{multirow}%
\usepackage{amsmath,amssymb,amsfonts}%
\usepackage{amsthm}%
\usepackage{mathrsfs}%
\usepackage[title]{appendix}%
\usepackage{xcolor}%
\usepackage{textcomp}%
\usepackage{manyfoot}%
\usepackage{booktabs}%
\usepackage{algorithm}%
\usepackage{algorithmicx}%
\usepackage{float}
\usepackage{makecell}
\usepackage{caption}
\usepackage{algpseudocode}%
\usepackage{needspace}
\usepackage[T1]{fontenc}       
\usepackage[utf8]{inputenc}%
\usepackage{listings}%
\usepackage{comment}

\lstdefinelanguage{PSUM-SysML2}{
    sensitive=true,
    morecomment=[l]{//},
    morecomment=[s]{/*}{*/},
    morekeywords={
        action,allocation,analysis,attribute,block,binding,calc,case,comment,concern,
        connection,constraint,def,doc,enum,flow,interface,item,metadata,objective,
        occurrence,package,part,port,ref,rendering,rep,requirement,snapshot,state,
        subject,succession,timeslice,transition,verification,view,viewpoint,
        about,abstract,accept,actor,after,alias,
        all,allocate,and,as,assert,assoc,assign,assume,at,
        bind,by,connect,crosses,decide,
        default,defined,dependency,derived,do,else,end,entry,event,
        exhibit,exit,expose,filter,first,for,fork,frame,from,hastype,if,
        implies,import,in,include,individual,inout,istype,join,language,
        loop,merge,message,nonunique,not,of,or,ordered,out,parallel,
        perform,private,protected,public,readonly,redefines,references,
        render,require,return,satisfy,send,specializes,stakeholder,
        standard,subsets,terminate,then,to,until,use,variant,variation,
        verify,via,when,while,xor
    },
    morekeywords=[2]{BeliefStatement,UncertaintyTopic,Uncertainty,IndeterminacySource,IndeterminacySpecification,
    Evidence,PersonalExperience,Effect,b_duration,u_pattern,u_reducibility,isr,mi,nd,ocr,con,tim,
    ale,epi,Fully reducible,PartiallyReducible,obj,subj,Random,emp,tpr,ied,ckn,measurement,m_precision,m_measurementError},
    morekeywords=[3]{vehicle.wheelDiameter},
    morekeywords=[4]{publicationPort, subscriptionPort},
    morestring=[b]",
}

\theoremstyle{thmstyleone}%
\theoremstyle{definition}%
\newtheorem{example}{Example}%

\theoremstyle{thmstyletwo}%

\theoremstyle{thmstylethree}%
\newtheorem{definition}{Definition}%

\newcommand{\yunyang}[1]{\textcolor{blue}{\textless Yunyang: #1\textgreater}}
\newcommand{\tao}[1]{\textcolor{orange}{\textless Tao: #1\textgreater}}

\usepackage{xspace}
\newcommand{\profile}{PSUM\text{-}SysMLv2\xspace}
\newcommand{\sysml}{SysML v2\xspace}
\newcommand{\modelcopilot}{ModelCopilot\xspace}

\usepackage{geometry}
\title{
Uncertainty Modeling for SysML v2
}

\author{Man Zhang}
\author{Yunyang Li}
\author{Tao Yue\thanks{Corresponding author}}

\affil{Beihang University}
\affil[ ]{\texttt{\{manzhang, liyunyang, yuetao\}@buaa.edu.cn}}

\date{}

\begin{document}

\maketitle

\begin{abstract}
Uncertainty is inherent in modern engineered systems, including cyber-physical systems, autonomous systems, and large-scale software-intensive infrastructures (such as microservice-based systems) operating in dynamic and partially observable environments. The recent publication of Precise Semantics for Uncertainty Modeling (PSUM) by the Object Management Group represents the first standardized specification for uncertainty modeling within the Model-Based Systems Engineering (MBSE) community, providing formally defined semantics for representing and reasoning about uncertainty in models. In parallel, the second version of Systems Modeling Language (\sysml) was released as the next-generation systems modeling language, offering improved semantic rigor and reusability, yet lacking native constructs aligned with PSUM for first-class uncertainty representation.
This paper proposes a systematic extension of \sysml that incorporates the PSUM metamodel into its modeling framework. The extension enables explicit specification of indeterminacy sources, structured characterization of uncertainties, and consistent propagation of uncertainty within system models, while preserving conformance with \sysml syntax and semantics. We validate the approach through seven case studies. Results demonstrate that the proposed extension (\profile) is expressive and applicable for uncertainty-aware MBSE, and potentially enables uncertainty and uncertainty propagation analyses.
\end{abstract}


\sloppy
\section{Introduction}\label{sec:introduction}
Modern software systems operate in environments characterized by variability, partial observability, and evolving operational conditions. Cyber-physical systems (CPS), autonomous platforms, and software-intensive infrastructures (such as microservice-based systems) must interact with uncertain physical processes, incomplete information, stochastic disturbances, and dynamic stakeholder requirements. Consequently, uncertainty is not an exception but a fundamental property of contemporary systems~\cite{troya2021uncertainty}. Ignoring or informally treating uncertainty at the early stage can lead to catestropic consequences.

In the field of software engineering, research on uncertainty dates back to the 1960s and 1970s, initially focusing on achieving determinism in software development processes~\cite{naur1969report}. Over time, it became widely recognized that epistemic uncertainty (e.g., from incomplete requirements, complex design decisions, and the unpredictability of user behaviors) permeates software system development lifecycle~\cite{laplante2005uncertainty}. With the widespread adoption of Artificial Intelligence (AI) and machine learning (ML) technologies, software systems are now subject to both epistemic and aleatoric uncertainties, arising from limitations in model knowledge, data noise, and the inherent variability of user interactions. As a result, uncertainty-aware software engineering has emerged as an increasingly important area of research.
Over the past decades, the modeling community has investigated various approaches to represent and reason about uncertainty, including probabilistic models~\cite{geng2017modeling,agli2016business,burgueno2018using,burgueno2019specifying}, interval-based specifications~\cite{vallecillo2016expressing}, fuzzy representations~\cite{sicilia2004extending,han2014extending, han2016fame, ma2007fuzzy, ma2011fuzzy, ma2012modeling}, and characterization and classification on Unified Modeling Langauge (UML) models~\cite{sedaghatbaf2018reliability,burgueno2019belief,burgueno2023dealing,zhang2019uncertainty}, etc. However, these approaches have often lacked standardized, formally defined semantics within the Model-Based Systems Engineering (MBSE) context. A significant milestone was recently achieved with the publication of Precise Semantics for Uncertainty Modeling (PSUM)~\cite{omg2024psum} by the Object Management Group (OMG). PSUM provides a metamodel-based foundation for representing uncertainty in a semantically precise manner, establishing the first standardized uncertainty modeling specification in the MBSE community.

At the same time, System Modeling Language (\sysml)~\cite{omg2025sysmlv2} was released as the next-generation systems modeling language. SysML v2 introduces a rigorously defined abstract syntax, improved semantic precision, and enhanced reusability mechanisms compared to the first version of SysML. These characteristics make it a promising foundation for integrating uncertainty modeling into mainstream systems engineering practice. Nevertheless, SysML v2 does not natively provide first-class constructs aligned with PSUM for explicitly representing uncertainty sources, uncertainty attributes, or uncertainty propagation mechanisms.

To this end, in this paper, we propose a systematic extension to SysML v2, named \profile, which incorporates the PSUM metamodel into SysML v2 and Kernal Modeling Language (KerML, the semantic and metamodeling foundation of SysML v2)~\cite{omg2025kerml}, to faciliate formal representation for uncertainty modeling through a coherent set of stereotypes, classes and constrained extensions.
To evaluate \profile, we conducted seven case studies collected from the literature.
Results demonstrate that \profile is expressive, semantically coherent, and
applicable for characterizing uncertainty systematically and explicitly in uncertainty-aware software/system engineering workflows.
We also observed that applying \profile potentially enables advanced uncertainty analyses, such as the identification of additional uncertainties, and uncertainty propagation analysis, though emprical evidence is required to confirm this.
We believe, by establishing a concrete bridge between PSUM and SysML v2, this work advances the integration of standardized uncertainty modeling into next-generation software and system engineering practice.
To effectively support uncertainty modeling, we plan to integrate \profile into our \modelcopilot\footnote{\label{foot:mc}\url{https://www.modelcopilot.org/model-copilot.html}}
platform and develop additional automated solutions to assist engineers in identifying, constructing, and evolving uncertainties within models.
We hope the community finds the profile useful and that it facilitates the implementation of uncertainty modeling across different platforms.
\profile is open sourced, and the models developed in our case studies are publicly available online\footnote{\url{https://github.com/WSE-Lab/PSUM-SysMLv2}}.

The remainder of this paper is organized as follows. Section~\ref{sec:background} introduces PSUM, SysML v2, and KerML, which are recently issued standards by OMG. In Section~\ref{sec:running examples}, we present the running examples (SysML v2 models) used to illustrate \profile. Section~\ref{sec:profile} presents the \profile\ profile itself. Section~\ref{sec:case_study} describes the conducted case studies. Related work is discussed in Section~\ref{sec:related_work}, and Section~\ref{sec:conclusion} concludes the paper.

\section{Background}\label{sec:background}

In this section, we discuss PSUM and \sysml, which are necessary to understand \profile.

\subsection{Precise Semantics for Uncertainty Modeling (PSUM)}\label{subsec:psum}
PSUM~\cite{omg2024psum} was built on U-Model, a conceptual model proposed by Zhang et al.~\cite{zhang2016understanding} (with tremandous contribution from Bran Selic), which defines uncertainty, its related concepts, and their relationships from the perspective of engineering large-scale software systems such as CPS and IoT. PSUM unifies the understanding of uncertainty in the software engineering domain and provides a standardized and reliable foundation for the development and application of uncertainty modeling.

\textit{PSUM::BeliefStatement} represents a representation of a belief specified in a specific language (e.g., \sysml) by \textit{PSUM::BeliefAgent}. A belief statement may be associated with one or more \textit{PSUM::Uncertainty}, which characterizes uncertainty from various aspects including uncertainty perspective (subjective or objective), pattern (e.g., periodic, persistent and sporadic), and effect (which itself can be an uncertainty), along with attributes such as kind (e.g., occurrence, content and time), nature (aleatory or epistemic), and reducibility (fully reducible, partially reducible and irreducible). \textit{PSUM::UncertaintyTopic} defines the specific aspect, parameter, or property to which the uncertainty applies, providing contextual scoping.

PSUM further distinguishes \textit{PSUM::IndeterminacySource}, which captures the origin of uncertainty, and is classified into: InsufficientResolution, MissingInfo, Non-determinism, Unclassified and Custom. These categories distinguish whether indeterminacy arises from insufficient precision, unavailable information, inherent or practical non-determinism, unspecified classification, or a user-defined source.

\textit{PSUM::Measurement} provides a generic mechanism to quantify belief, uncertainty, risk, and indeterminacy source, which may comprise measurable features such as accuracy, sensitivity, measurement error, precision, or degree. Each feature is associated with the Structured Metrics Metamodel (SMM) standard~\cite{omg2025smm} at OMG.

In summary, a belief statement anchors an assertion; uncertainty reflects the degree of confidence or lack of knowledge associated with the belief statement; indeterminacy source explains the origin of the uncertainty; uncertainty topic scopes its domain; and measurement formalizes its quantification. Together, these constructs establish a structured and extensible foundation for uncertainty-aware system modeling.
In the rest of this paper, specific PSUM elements are introduced as they arise.

\subsection{Systems Modeling Language Version 2 (SysML v2)}\label{subsec:SysML v2}
\sysml~\cite{omg2025sysmlv2} addresses limitations of SysML v1 by providing a cleaner and more rigorous semantic foundation. Unlike SysML v1, which was defined as a profile of UML, \sysml is defined using KerML~\cite{omg2025kerml}, a formal modeling kernel that establishes a unified semantic basis for structural and behavioral modeling.

\sysml introduces a systematic distinction between \textit{SysML::Definition} and \textit{SysML::Usage}, enabling a clear separation between types and their contextual applications. It further supports reuse and refinement through other three specialization mechanisms: subclassification, subsetting, and redefinition.
In addition, \sysml provides a fully specified textual notation alongside graphical representations. Note that, in this paper, we focus on textual notations of \sysml.
More information about \sysml, one can refer to its OMG specification~\cite{omg2025sysmlv2}.
In the remainder of this paper, specific modeling constructs and notations are introduced as they arise.

\section{Running Examples}\label{sec:running examples}

The three running examples are:
(1) Adaptive Cruise Control (ACC),
(2) Interaction Sequencing (IS), and
(3) Vehicle Fuel Economy Analysis (VFEA).
We present the partial examples in \sysml textual notations extended with textual notations of \profile.
Specifically, we want to clarify that syntax extensions introduced by the \profile profile are highlighted in blue; the red color indicates uncertain elements derived via specialization relationships of SysML v2 (e.g., uncertain usages inherited from explicitly modelled uncertain definitions); and orange text denotes indeterminacy sources inherited via specializations. We also follow a commonly-seen modeling conventions of \sysml: the first letters of the names of definitions are capitalized, while those of usages begin with lowercase letters.


\subsection{The ACC Running Example}
ACC involves three key modules: perception, control, and decision-making under uncertainty, aiming to assist vehicles in maintaining a safe following distance while adhering to speed limits. Specifically, ACC's perception module measures the relative distance, speed, and acceleration of the leading vehicle via sensors including lasers, radars and cameras~\cite{ali2024adaptive}.

As shown in Example~\ref{example:acc}, structurally, to maintain a safe distance from other vehicles, ACC perceives its surroundings through a set of sensors, which are modelled as part usages of part definition \textit{ACC}: \emph{radars}, \textit{lidars} and \textit{cameras}. Their definitions \emph{Radar}, \emph{Lidar}, and \emph{Camera} are defined in Lines~\ref{line:def_radar}-\ref{line:end_camera}, accordingly. These part definitions are specialized from part definition \emph{Sensor} which is characterized with attributes such as horizontal and vertical fields of a view (Lines~\ref{line:def_sensor}-\ref{line:end_sensor}).
The ACC behavior (partial) is essentially exhibited as two parallel state machines: \emph{perceptionLayerState} and \emph{decisionLayerState} (Lines~\ref{line:state_perception} and~\ref{line:state_decision}). ACC starts from state \emph{idle} and then initiates perception by transitioning to state \emph{perceiving} (Lines~\ref{line:percep_entry}-\ref{line:trans_start_percep}). Within this state, action \emph{perceive} for continuously measuring key parameters of the leading vehicle is executed, and the sensed data is sent to the decision layer through \emph{PerceptionSignal} (Line~\ref{line:in_percep_then}). In the state machine of the decision layer, ACC is initially in state \emph{waitingForSignal} (Line~\ref{line:state_wait_signal}). Upon receiving \emph{PerceptionSignal}, it transits to the \emph{deciding} state (Lines~\ref{line:state_deciding}-\ref{line:end_deciding}), where control decision (e.g., acceleration of deceleration commands) are computed.

\begin{example}
    Adaptive Cruise Control (ACC)
    \label{example:acc}
    \begin{lstlisting}
package StructuralModel { (*@\label{line:pkg_structural}@*)
	part def `ACC' { (*@\label{line:def_acc_system}@*)
		// Radar could be physically blocked or disabled with IndeterminayNature being Non-determinism (`nd')
		«IndeterminacySource<nd>» part radars defined by Radar[*] { (*@\label{line:part_radar}@*)
			«IndeterminacySpecification» constraint radarBlocked { (*@\label{line:const_radar_blocked}@*)
				isBlocked == true (*@\label{line:expr_blocked_true}@*) } (*@\label{line:end_const_blocked}@*)
			«IndeterminacySpecification» constraint radarNotBlocked { (*@\label{line:const_radar_not_blocked}@*)
				isBlocked == false (*@\label{line:expr_blocked_false}@*) } (*@\label{line:end_const_not_blocked}@*)
		} (*@\label{line:end_part_radar}@*)
		// Data from lasers and cameras could lack sufficient precision or fidelity to faithfully capture observed phenomenon, with IndeterminayNature being InsufficientResolution (`isr')
		«IndeterminacySource<isr>» part lidars defined by Lidar[*]; (*@\label{line:part_laser}@*)
		«IndeterminacySource<isr>» part cameras defined by Camera[*]; (*@\label{line:part_camera}@*)
	} (*@\label{line:end_acc_system}@*)
	part def Sensor { (*@\label{line:def_sensor}@*)
		// Horizontal and vertical fields of a view (in degrees)
		attribute fovHorizontal defined by ScalarValues::Real; (*@\label{line:attr_fov_h}@*)
		attribute fovVertical defined by ScalarValues::Real; (*@\label{line:attr_fov_v}@*)
	} (*@\label{line:end_sensor}@*)
	part def Radar specializes Sensor { (*@\label{line:def_radar}@*)
		// Maximum detection range (in meters)
		attribute range defined by ScalarValues::Real; (*@\label{line:attr_radar_range}@*)
		// Indicates whether the radar's field of view is obstructed
		attribute isBlocked defined by ScalarValues::Boolean; (*@\label{line:attr_is_blocked}@*)
	} (*@\label{line:end_radar}@*)
	part def Lidar specializes Sensor; (*@\label{line:def_lidar}@*)
	part def Camera specializes Sensor { (*@\label{line:def_camera}@*)
		// Image resolution (in megapixels, MP)
		attribute resolution defined by ScalarValues::Real; (*@\label{line:attr_camera_res}@*)
	} (*@\label{line:end_camera}@*)
} (*@\label{line:end_pkg_structural}@*)
package BehavioralModel { (*@\label{line:pkg_behavioral}@*)
	private import StructuralModel::*;
	private import SignalDefinition::*;
	state def ACCState { (*@\label{line:def_acc_state}@*)
		part acc: ACC;
		entry; (*@\label{line:acc_entry}@*)
		then idle; (*@\label{line:acc_then_idle}@*)sub
		state idle; (*@\label{line:state_idle}@*)
		state ready; (*@\label{line:state_ready}@*)
		state accOn parallel { (*@\label{line:state_acc_on}@*)
			state perceptionLayerState { (*@\label{line:state_perception}@*)
				entry; (*@\label{line:percep_entry}@*)
				then idle; (*@\label{line:percep_then_idle}@*)
				state idle; (*@\label{line:percep_idle}@*)
				// Perceive the leading vehicle's current distance, speed, acceleration, etc.
				state perceiving {do action perceive;}(*@\label{line:action_perceive}@*)
				transition startPerception first idle then perceiving; (*@\label{line:trans_start_percep}@*)
				transition inPerception first perceiving
				do send PerceptionSignal() then perceiving; (*@\label{line:in_percep_then}@*)
			}
			«BeliefStatement» state decisionLayerState  { (*@\label{line:state_decision}@*)
				b_duration = 30 [SI::day]; (*@\label{line:belief_duration}@*)
				entry; (*@\label{line:decision_entry}@*)
				then waitingForSignal; (*@\label{line:decision_then_wait}@*)
				state waitingForSignal; (*@\label{line:state_wait_signal}@*)
				state deciding { (*@\label{line:state_deciding}@*)
					// Make control decision based on the received signal from the perception layer.
					«Effect» do action `decide'; (*@\label{line:action_decide}@*)} (*@\label{line:end_deciding}@*)
				then waitingForSignal; (*@\label{line:deciding_then_wait}@*)
				// The radar is not blocked and the decision layer can receive PerceptionSignal.
				«Uncertainty<ocr, epi, subj>» transition startDeciding (*@\label{line:trans_start_deciding}@*)
				first waitingForSignal (*@\label{line:start_deciding_first}@*)
				accept PerceptionSignal (*@\label{line:start_deciding_accept}@*)
				then deciding { (*@\label{line:start_deciding_then}@*)
					u_reducibility = PartiallyReducible; (*@\label{line:u_reducibility_1}@*)
					u_pattern = Random; (*@\label{line:u_pattern_1}@*)
					// `::>' represents references (reserence subsetting)
					«IndeterminacySpecification» ref ::> acc.radar.radarNotBlocked; (*@\label{line:ref_not_blocked}@*)
					«Effect» ref ::> deciding. `decide'; (*@\label{line:ref_effect_decide}@*)
				} (*@\label{line:end_trans_start_deciding}@*)
				// The radar is blocked, and the decision layer cannot receive PerceptionSignal
				«Uncertainty<ocr, epi, subj>» transition failToStartDeciding (*@\label{line:trans_fail_deciding}@*)
				first waitingForSignal (*@\label{line:fail_deciding_first}@*)
				accept PerceptionSignal (*@\label{line:fail_deciding_accept}@*)
				then waitingForSignal { (*@\label{line:fail_deciding_then}@*)
					u_reducibility = PartiallyReducible; (*@\label{line:u_reducibility_2}@*)
					u_pattern = Random; (*@\label{line:u_pattern_2}@*)
					«IndeterminacySpecification» ref ::> acc.radar.radarBlocked; (*@\label{line:ref_blocked}@*)
				} (*@\label{line:end_trans_fail_deciding}@*)
				metadata collisionRisk defined by RiskMetadata::Risk about failToStartDeciding{ (*@\label{line:risk_collision}@*)
					totalRisk {impact = RiskMetadata::LevelEnum::high; (*@\label{line:risk_impact}@*)}}} (*@\label{line:end_decision_layer}@*)
		} (*@\label{line:end_acc_on}@*)
		transition setACC first idle then ready; (*@\label{line:trans_set_acc}@*)
		transition accError first idle then error; (*@\label{line:trans_acc_error}@*)
		transition turnOnACC first ready accept ACCTurnOnSignal then accOn; (*@\label{line:trans_turn_on}@*)
		transition turnOffACC first accOn accept ACCTurnOffSignal then idle; (*@\label{line:trans_turn_off}@*)
	} (*@\label{line:end_acc_state}@*)
} (*@\label{line:end_pkg_behavioral}@*)
package SignalDefinition { (*@\label{line:pkg_signal}@*)
	private import BehavioralModel::ACCState::*; (*@\label{line:import_acc_state}@*)
	«UncertaintyTopic» item def PerceptionSignal { (*@\label{line:def_percep_signal}@*)
		«Uncertainty» ref ::> accOn.decisionLayerState.startDeciding; (*@\label{line:ref_u_start}@*)
		«Uncertainty» ref ::> accOn.decisionLayerState.failToStartDeciding; (*@\label{line:ref_u_fail}@*)
	} (*@\label{line:end_percep_signal}@*)
	item def ACCTurnOnSignal; (*@\label{line:def_on_signal}@*)
	item def ACCTurnOffSignal; (*@\label{line:def_off_signal}@*)
	// Further details omitted for brevity
} (*@\label{line:end_pkg_signal}@*)
    \end{lstlisting}
\end{example}

\subsection{The IS Running Example}

The SysML model of IS (partial) describes a publish-subscribe communication scenario involving three roles: \emph{producer}, \emph{server}, and \emph{consumer}.
Structurally, as shown in Example~\ref{example:is}, part \emph{producer} is equipped with a publication port to send publications (Line~\ref{line:port_prod_pub}), part \emph{consumer} has a subscription port to subscribe its interest in certain topics (Line~\ref{line:port_cons_sub}), and part \emph{server} exposes both a publication port and a subscription port to mediates the interaction by managing subscriptions and delivering publications between producers and consumers (Lines~\ref{line:port_srv_pub}-\ref{line:port_srv_sub}).
From a behavioral perspective, part \emph{consumer} starts by performing a \emph{subscribe} action to register its topic of interest with \emph{server} (Lines~\ref{line:act_subscribe}-\ref{line:end_act_subscribe}). Part \emph{server} waits for subscriptions in its initial state (\emph{waitForSubscription}) and, upon receiving a subscription request, transits to \emph{waitForPublication} (Lines~\ref{line:trans_subscribing}-\ref{line:end_trans_subscribing}). Meanwhile, \emph{producer} executes action \emph{publish} to send a publication on the specified topic (Lines~\ref{line:act_publish}-\ref{line:end_act_publish}). When part \emph{server} receives this publication and matches it with the active subscription, it triggers a \emph{delivering} transition to deliver the publication to \emph{consumer} (Lines~\ref{line:trans_delivering}-\ref{line:end_trans_delivering}). These three roles communicate with each other via signals using the ports defined in the structural model.

\begin{example}
    Interaction Sequencing system
    \label{example:is}
    \begin{lstlisting}
package SignalDefinitions { (*@\label{line:pkg_signal_def}@*)
	private import Configuration::*; (*@\label{line:import_config}@*)
	«UncertaintyTopic» item def Subscribe { (*@\label{line:item_subscribe}@*)
		attribute topic defined by ScalarValues::String; (*@\label{line:attr_sub_topic}@*)
		ref part subscriber; (*@\label{line:ref_subscriber}@*)
		«Uncertainty» ref ::> server.serverBehavior.subscribing; (*@\label{line:u_ref_sub_server}@*)
		«Uncertainty» ref ::> consumer.consumerBehavior.subscribe; (*@\label{line:u_ref_sub_consumer}@*)
	} (*@\label{line:end_item_subscribe}@*)
	«UncertaintyTopic» item def Publish { (*@\label{line:item_publish}@*)
		attribute topic defined by ScalarValues::String; (*@\label{line:attr_pub_topic}@*)
		ref publication; (*@\label{line:ref_publication}@*)
		«Uncertainty» ref ::> producer.producerBehavior.publish; (*@\label{line:u_ref_pub_producer}@*)
		«Uncertainty» ref ::> server.serverBehavior.delivering; (*@\label{line:u_ref_pub_server}@*)
	} (*@\label{line:end_item_publish}@*)
	«UncertaintyTopic» item def Deliver { (*@\label{line:item_deliver}@*)
		ref publication; (*@\label{line:ref_del_pub}@*)
		«Uncertainty» ref ::> server.serverBehavior.delivering; (*@\label{line:u_ref_del_server}@*)
		«Uncertainty» ref ::> consumer.consumerBehavior.delivery; (*@\label{line:u_ref_del_consumer}@*)
	} (*@\label{line:end_item_deliver}@*)
} (*@\label{line:end_pkg_signal_def}@*)
package Configuration { (*@\label{line:pkg_config}@*)
	private import SignalDefinitions::*;
	constraint def Operational { (*@\label{line:const_def_op}@*)
		in status defined by ScalarValues::Boolean; (*@\label{line:in_status_op}@*)
		status == true (*@\label{line:expr_op_true}@*)}
	constraint def NotOperational { (*@\label{line:const_def_nop}@*)
		in status defined by ScalarValues::Boolean; (*@\label{line:in_status_nop}@*)
		status == false (*@\label{line:expr_nop_false}@*)}
	// The behavior of PublicationPort and SubscriptionPort is Non-determinism (`nd')
	«IndeterminacySource<nd>» port def PublicationPort { (*@\label{line:port_def_pub}@*)
		attribute operationalStatus defined by ScalarValues::Boolean; (*@\label{line:attr_pub_status}@*)
		«IndeterminacySpecification» constraint publicationPortOperational defined by Operational { (*@\label{line:spec_pub_op}@*)
			in status = operationalStatus; (*@\label{line:map_pub_op}@*)}
		«IndeterminacySpecification» constraint publicationPortNotOperational defined by NotOperational { (*@\label{line:spec_pub_nop}@*)
			in status = operationalStatus; (*@\label{line:map_pub_nop}@*)}
	} (*@\label{line:end_port_def_pub}@*)
	«IndeterminacySource<nd>» port def SubscriptionPort { (*@\label{line:port_def_sub}@*)
		attribute operationalStatus defined by ScalarValues::Boolean; (*@\label{line:attr_sub_status}@*)
		«IndeterminacySpecification» constraint subscriptionPortOperational defined by Operational { (*@\label{line:spec_sub_op}@*)
			in status = operationalStatus; (*@\label{line:map_sub_op}@*)}
		«IndeterminacySpecification» constraint subscriptionPortNotOperational defined by NotOperational { (*@\label{line:spec_sub_nop}@*)
			in status = operationalStatus; (*@\label{line:map_sub_nop}@*)}
	} (*@\label{line:end_port_def_sub}@*)
	part producer[1] { (*@\label{line:part_producer}@*)
		attribute someTopic defined by ScalarValues::String; (*@\label{line:attr_prod_topic}@*)
		private item somePublication; (*@\label{line:item_prod_pub}@*)
		port publicationPort defined by ~PublicationPort; (*@\label{line:port_prod_pub}@*)
		«BeliefStatement» perform action producerBehavior { (*@\label{line:act_prod_behavior}@*)
			«Uncertainty<ocr, epi, subj>» action publish send Publish(someTopic, somePublication) via publicationPort { (*@\label{line:act_publish}@*)
				// `::>' represents for references (reference subsetting in SysML v2)
				«IndeterminacySpecification» ref ::> publicationPort.publicationPortOperational; (*@\label{line:ref_pub_port_op}@*)
				«Effect» ref :>> server.serverBehavior.delivering; (*@\label{line:ref_eff_del}@*)
			} (*@\label{line:end_act_publish}@*)
		} (*@\label{line:end_act_prod_behavior}@*)
	} (*@\label{line:end_part_producer}@*)
	part server[1] { (*@\label{line:part_server}@*)
		port publicationPort defined by PublicationPort; (*@\label{line:port_srv_pub}@*)
		port subscriptionPort defined by SubscriptionPort; (*@\label{line:port_srv_sub}@*)
		«BeliefStatement» exhibit state serverBehavior { (*@\label{line:state_srv_behavior}@*)
			entry; then waitForSubscription; (*@\label{line:srv_entry}@*)
			state waitForSubscription; (*@\label{line:state_wait_sub}@*)
			state waitForPublication; (*@\label{line:state_wait_pub}@*)
			«Uncertainty<ocr, epi, subj>, Effect» transition subscribing (*@\label{line:trans_subscribing}@*)
			first waitForSubscription (*@\label{line:sub_first}@*)
			accept sub defined by Subscribe via subscriptionPort (*@\label{line:sub_accept}@*)
			then waitForPublication { (*@\label{line:sub_then}@*)
				«IndeterminacySpecification» ref ::> consumer.subscriptionPort.subscriptionPortOperational; (*@\label{line:ref_cons_port_op}@*)
				«IndeterminacySpecification» ref ::> subscriptionPort.subscriptionPortOperational; (*@\label{line:ref_srv_pub_op}@*)
			} (*@\label{line:end_trans_subscribing}@*)
			«Uncertainty<ocr, epi, subj>, Effect» transition delivering (*@\label{line:trans_delivering}@*)
			first waitForPublication (*@\label{line:del_first}@*)
			accept pub defined by Publish via publicationPort (*@\label{line:del_accept}@*)
			if pub.topic == subscribing.sub.topic (*@\label{line:del_guard}@*)
			do send Deliver(pub.publication) to subscribing.sub.subscriber (*@\label{line:del_do}@*)
			then waitForPublication { (*@\label{line:del_then}@*)
				«IndeterminacySpecification» ref ::> producer.publicationPort.publicationPortOperational; (*@\label{line:ref_prod_port_op}@*)
				«IndeterminacySpecification» ref ::> publicationPort.publicationPortOperational; (*@\label{line:ref_srv_pub_op2}@*)
				«Effect» ref ::> consumer.consumerBehavior.delivery; (*@\label{line:ref_eff_cons_del}@*)
			} (*@\label{line:end_trans_delivering}@*)
		} (*@\label{line:end_state_srv_behavior}@*)
	} (*@\label{line:end_part_server}@*)
	part consumer[1] { (*@\label{line:part_consumer}@*)
		attribute myTopic defined by ScalarValues::String; (*@\label{line:attr_cons_topic}@*)
		port subscriptionPort defined by ~SubscriptionPort; (*@\label{line:port_cons_sub}@*)
		«BeliefStatement» perform action consumerBehavior { (*@\label{line:act_cons_behavior}@*)
			«Uncertainty<ocr, epi, subj>» action subscribe send Subscribe(myTopic, consumer) via subscriptionPort to server{ (*@\label{line:act_subscribe}@*)
				«IndeterminacySpecification» ref ::> subscriptionPort.subscriptionPortOperational; (*@\label{line:ref_cons_sub_op}@*)
				«Effect» ref ::> server.serverBehavior.subscribing; (*@\label{line:ref_eff_srv_sub}@*)
			} (*@\label{line:end_act_subscribe}@*)
			then «Effect» action delivery accept Deliver via consumer { (*@\label{line:act_delivery}@*)
				// Futher details omitted for brevity
			}
		}
	}
}
    \end{lstlisting}
\end{example}

\subsection{The VFEA Running Example}
The VFEA running example models a vehicle fuel economy analysis case. As shown in Example~\ref{exampe:vfea}, structurally, part definition \emph{Vehicle} is characterized with attributes \emph{mass}, \emph{cargoMass}, \emph{wheelDiameter}, and \emph{driveTrainEfficiency}, and has port \emph{fuelInPort} (Lines~\ref{line:def_vehicle}-\ref{line:end_vehicle}). The core of this model lies in analysis definition \emph{FuelEconomyAnalysis}, which evaluates whether a given part usage of \emph{Vehicle} satisfies the fuel economy requirement under specific driving scenarios (Lines~\ref{line:def_analysis}-\ref{line:end_analysis}).

\begin{example}
    Vehicle Fuel Economy Analysis system
    \label{exampe:vfea}
    \begin{lstlisting}
package VehicleModel { (*@\label{line:pkg_vehicle}@*)
	item def Fuel; (*@\label{line:item_fuel}@*)
	port def FuelPort { (*@\label{line:port_def_fuel}@*)
		out item fuel defined by Fuel; (*@\label{line:port_fuel_out}@*)
	} (*@\label{line:end_port_def_fuel}@*)
	«BeliefStatement» part def Vehicle { (*@\label{line:def_vehicle}@*)
		attribute mass defined by ISQ::MassValue; (*@\label{line:attr_mass}@*)
		attribute cargoMass defined by ISQ::MassValue; (*@\label{line:attr_cargo_mass}@*)
		// The assumed vehicle parameters constitute a content uncertainty, since precise constant values are used to represent physical quantities that vary continuously due to manufacturing tolerances, wear, and operating conditions.
		«Uncertainty<con, ale, subj>» attribute wheelDiameter defined by ISQ::LengthValue { (*@\label{line:attr_wheel_dia}@*)
			measurement { (*@\label{line:meas_start}@*)
				m_measurementError = 1.5[`%']; (*@\label{line:meas_error}@*)
			} (*@\label{line:meas_end}@*)
		} (*@\label{line:end_wheel_dia}@*)
		attribute driveTrainEfficiency defined by ScalarValues::Real; (*@\label{line:attr_efficiency}@*)
		port fuelInPort defined by ~FuelPort; (*@\label{line:port_fuel_in}@*)
	} (*@\label{line:end_vehicle}@*)
} (*@\label{line:end_pkg_vehicle}@*)
package FuelEconomyAnalysisModel { (*@\label{line:pkg_analysis}@*)
	private import VehicleModel::*; (*@\label{line:import_vehicle}@*)
	analysis def FuelEconomyAnalysis { (*@\label{line:def_analysis}@*)
		subject vehicle defined by Vehicle; (*@\label{line:subj_vehicle}@*)
		in attribute scenario; (*@\label{line:in_scenario}@*)
		in requirement fuelEconomyRequirement; (*@\label{line:in_req}@*)
		return calculatedFuelEconomy; (*@\label{line:ret_economy}@*)
		objective fuelEconomyAnalysisObjective { (*@\label{line:obj_analysis}@*)
			doc /* The objective of this analysis is to determine whether the current vehicle design configuration can satisfy the fuel economy requirement. */ (*@\label{line:doc_start}@*)
			assume constraint fuelEconomyAnalysisAssumedConstraint{ (*@\label{line:const_assume}@*)
				// vehicle.wheelDiameter == 33 [`inch'] & (original assumed value) (*@\label{line:expr_dia_original}@*)
				vehicle.wheelDiameter >= 32.505 [`inch'] & (*@\label{line:expr_dia_min}@*)
				vehicle.wheelDiameter <= 33.495 [`inch'] & (*@\label{line:expr_dia_max}@*)
				vehicle.driveTrainEfficiency == 0.4 (*@\label{line:expr_eff_val}@*)
			} (*@\label{line:end_const_assume}@*)
			require fuelEconomyRequirement; (*@\label{line:req_economy}@*)
		} (*@\label{line:end_objective}@*)
		// Further details omitted for brevity
	} (*@\label{line:end_analysis}@*)
}
    \end{lstlisting}
\end{example}


\section{The \profile Profile}\label{sec:profile}

The \profile profile extends SysML v2 and KerML with the PSUM metamodel, to enable explicit representation of uncertainty, its related concepts as well as their measurements.
Based on the structure of PSUM, we organize the \profile profile into three sub-profiles: Belief, Uncertainty, and Measurement.
The Belief profile captures beliefs held by belief agents, where a belief is an implicit conceptualization of phenomena or notions and is made explicit through one or more belief statements (Section~\ref{subsec:belief_profile}).
The Uncertainty profile formulates uncertainty as the state of deficiency in information or knowledge regarding an uncertainty topic and characterizes uncertainty with properties including uncertainty kind (e.g., content, time, occurrence), nature (aleatory or epistemic), reducibility level, and perspective (subjective or objective), and further described by characteristics like pattern and effect (Section~\ref{subsec:uncertainty_profile}).
The Measurement profile provides ways to quantify measurable PSUM elements (e.g., belief, uncertainty), using features like degree, precision, and accuracy, thereby enabling the integration of quantitative reasoning into uncertainty modeling (Section~\ref{subsec:measurement_profile}).

The complete mappings and extensions between stereotypes or classes in \profile and metaclasses in KerML or SysML v2 is provided in Table~\ref{tab:PSUM-SysML v2}.
In the remainder of this section, we discuss several example metaclasses to which a stereotype can be applied.

\begin{table}[!h]
    \centering
    \resizebox{\textwidth}{!}{%
        \begin{minipage}{\textwidth}
            \caption{Extensions of \profile stereotypes/classes to KerML/\sysml metaclasses}
            \label{tab:PSUM-SysML v2}
             \resizebox{.99\linewidth}{!}{%
            \begin{tabular}{@{}lll@{}}
                \toprule
                Profile & Stereotype/class in \profile & Metaclass in KerML/\sysml \\
                \midrule
                Belief
                & \guillemotleft BeliefStatement\guillemotright & KerML::Element \\
                & \guillemotleft IndeterminacySource\guillemotright &
                \makecell[l]{SysML::AttributeDefinition, SysML::OccurrenceDefinition, \\
                SysML::AttributeUsage, SysML::OccurrenceUsage} \\
                & \guillemotleft IndeterminacySpecification\guillemotright & SysML::ConstraintUsage \\
                \midrule
                Uncertainty
                & \guillemotleft Uncertainty\guillemotright &
                \makecell[l]{SysML::AttributeDefinition, SysML::OccurrenceDefinition, \\
                SysML::AttributeUsage, SysML::OccurrenceUsage} \\
                & \guillemotleft UncertaintyTopic\guillemotright &
                \makecell[l]{SysML::AttributeDefinition, SysML::OccurrenceDefinition, \\
                SysML::AttributeUsage, SysML::OccurrenceUsage} \\
                & \guillemotleft Effect\guillemotright &
                \makecell[l]{SysML::AttributeDefinition, SysML::OccurrenceDefinition, \\
                SysML::AttributeUsage, SysML::OccurrenceUsage} \\
                & Risk & SysML::MetadataUsage\footnotemark[1] \\
                \midrule
                Measurement
                & Accuracy & KerML::Expression\footnotemark[2] \\
                & Sensitivity & KerML::Expression\footnotemark[2] \\
                & MeasurementError & KerML::Expression\footnotemark[2] \\
                & Precision & KerML::Expression\footnotemark[2] \\
                & Degree & KerML::Expression\footnotemark[2] \\
                \bottomrule
            \end{tabular}
        }
            \footnotetext[1]{Realized through SysML::MetadataUsage instance, which is defined by MetadataDefinition RiskMetadata::Risk provided in standard libraries.}
            \footnotetext[2]{Associated with KerML::Expression through measurement.}
        \end{minipage}%
    }
\end{table}

\subsection{Dealing with Specialization Relationships of SysML v2}\label{subsec:defUsagePair}
In \profile, stereotypes are applied systematically following the definition-usage mechanism, and more broadly, the four key specialization relationships: subclassification between definitions, subsetting between usages, redefinition between usages, and definition between definitions and usages.

Since definition elements capture intrinsic properties and features, while usage elements represent their instantiation in specific contexts, stereotypes of \profile can be applied to a definition, allowing all usages defined by it to inherit the properties and avoid redundant specifications, or directly to a usage to model context-specific or instance-level variations. Subclassification between definitions enables specialized definitions to inherit features and stereotypes from general definitions. Subsetting between usages restricts the values of a usage while inheriting its stereotypes. Redefinition between usages allows a redefining usage to replace or override features of the usage it redefines with optional stereotype inheritance or override.
Methodologically, stereotyping definitions is primarily for reuse, while stereotyping usages captures contextual nuances. By combining definition-usage separation with systematic handling of specialization, \profile ensures consistent, non-redundant stereotype application and supports modeling both type-level intrinsic properties and instance-level contextual variations.

\subsection{The Belief Profile}\label{subsec:belief_profile}
\begin{figure}[H]
    \centering
    \includegraphics[width=\textwidth]{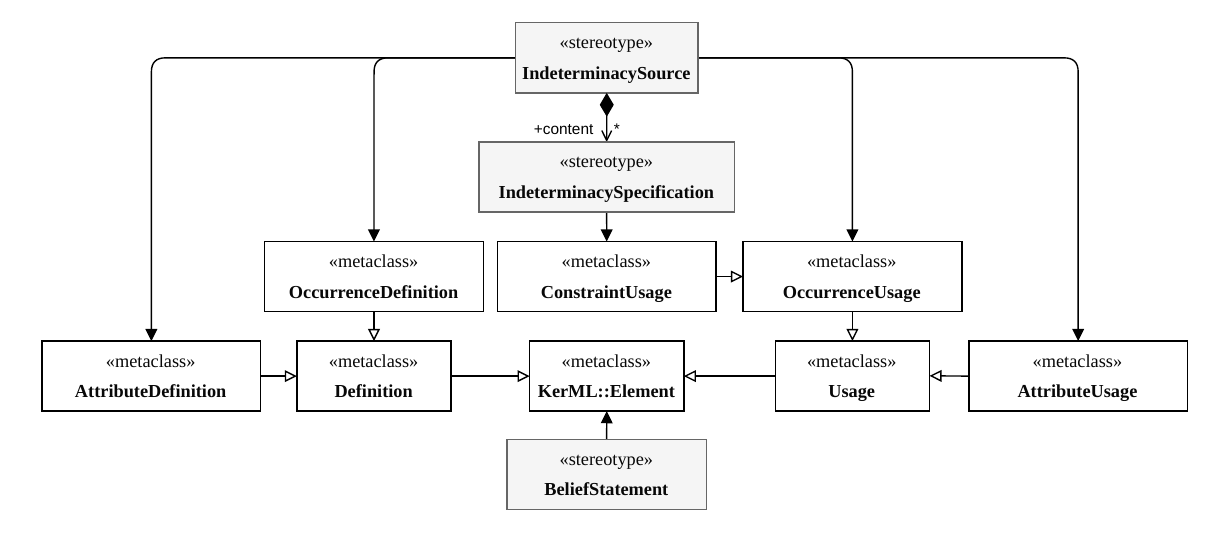}
    \caption{The belief profile of \profile. The classes in grey are stereotypes while those in white are either KerML or SysML v2 metaclasses.}\label{fig:belief_profile}
\end{figure}

The belief profile of \profile aims to implement the PSUM belief metamodel package.
When modeling with SysML v2, we do not need to explicitly capture \textit{PSUM::BeliefAgent}, because, its role is explicitly played by the modeler who creates the SysML v2 model itself. It is also not needed to capture \textit{PSUM::Belief}, as our context of interest is only about modeling with SysML v2 and we do not need to capture how the same belief can be modeled in different ways.
Hence, as shown in Figure~\ref{fig:belief_profile}, we define three stereotypes in this profile:
\guillemotleft BeliefStatement\guillemotright,
\guillemotleft IndeterminacySource\guillemotright, and
\guillemotleft IndeterminacySpecification\guillemotright, which are discussed in detail below.

\subsubsection{Belief Statement}\label{subsubsec:beliefstatement}

To model uncertainty within a system, we first need to specify belief statements, to signify which SysML model elements represent beliefs of the modeler(s).
In \profile, \textit{PSUM::BeliefStatement} is designed as a stereotype (\guillemotleft BeliefStatement\guillemotright) that can be applied to any \sysml elements, and hence extends \textit{KerML::Element}, as all \sysml model elements are actually specializations of \textit{KerML::Element} (Section~\ref{subsec:SysML v2}).  Doing so allows us to denote any \sysml model element (e.g., states and actions) as belief statements, which consequently are associated to uncertainties and indeterminacy sources (Section~\ref{subsec:uncertainty_profile}).

Below, we first define the stereotype, then illustrate its application in representative structural and behavioral modeling contexts.

\begin{definition}
    \textbf{\guillemotleft BeliefStatement\guillemotright}
    \guillemotleft BeliefStatement\guillemotright\ can be applied to any SysML v2 model element to denote that the element is a concrete representation of a belief of a modeler. A belief statement can be further associated to uncertainties and indeterminacy sources if needed as specified in PSUM.
\end{definition}

\textbf{(a) Part.}
In SysML v2, a part represents a modular unit of a system structure, including the system itself, its components, or external entities, which can be modeled as \textit{SysML::PartDefinition} and \textit{SysML::PartUsage}~\cite{OMG-SysML-2.0}. When the system is modeled as a composite part, its part usages themselves may also be composite and can own features such as attributes. Naturally, \guillemotleft BeliefStatement\guillemotright\ may be applied to parts to indicate that the composite structure of the part is subject to uncertainties.
As shown in Example~\ref{exampe:vfea}), part definition \emph{Vehicle}, which defines the analysis subject of \emph{FuelEconomyAnalysis}, is stereotyped with \guillemotleft BeliefStatement\guillemotright\ (Lines~\ref{line:def_vehicle}), implying that its attributes might be subject to uncertainty. For instance, attribute \emph{wheelDiameter} in Line~\ref{line:attr_wheel_dia} is uncertain due to manufacturing tolerances, wear or operating conditions (see Section~\ref{subsubsec:uncertainty} for more details).

\textbf{(b) State.}
State machines are commonly used to model discrete behavioral dynamics, where system behavior is represented as states and transitions between them. In SysML v2, \textit{SysML::StateDefinition} and \textit{SysML::StateUsage} are used to define states and their usages, by the capability of capturing  which may compose substates, actions, transitions, constraints (state invariant), etc.~\cite{OMG-SysML-2.0}.
%
As shown in Example~\ref{example:acc}, state definition \emph{ACCState} consists of states \emph{idle}, \emph{ready}, and \emph{accOn} (Lines~\ref{line:state_idle}-\ref{line:state_acc_on}). Especially, \emph{accOn} is a parallel state with substates \emph{perceptionLayerState}, and \emph{decisionLayerState}, on which we apply \guillemotleft BeliefStatement\guillemotright\ (Line~\ref{line:state_decision}), indicating that the state-based behavior of the decision layer in ACC may contain uncertainties (e.g., the system may fail to transit into the state of making control decisions when the perception signal is indeterminate due to sensor noise or failure) (see Section~\ref{subsubsec:uncertainty} for more details).
Since an explicit \textit{PSUM::Belief} is not introduced, the duration, representing the time period during which the belief is considered valid~\cite{omg2024psum}, can be specified directly within the belief statement (Lines~\ref{line:belief_duration}) when needed. For instance, the duration \emph{30 [SI::day]} is expressed using \textit{KerML::Expression}, leveraging time quantities defined in the standard library package of SysML v2.

\textbf{(c) Action.}
Actions also play a central role in modeling system behavior. As defined in the SysML v2 specification, an action is a kind of occurrences that can coordinate the performance of other actions and generate effects on system components involved in its execution~\cite{omg2025sysmlv2}. In SysML v2, actions are specified with \textit{SysML::ActionDefinition} and \textit{SysML::ActionUsage}. When an action is declared as a belief statement, its parameters or subactions may be specified as uncertain elements, indicating that uncertainties exist in the execution of the action.
As shown in Example~\ref{example:is}, action usage \emph{producerBehavior} performed by part \emph{producer} and \emph{consumerBehavior} performed by part \emph{consumer} are stereotyped with \guillemotleft BeliefStatement\guillemotright\ (Lines~\ref{line:act_prod_behavior} and~\ref{line:act_cons_behavior}). Since the actual execution of \emph{producerBehavior} and \emph{consumerBehavior} depends on sub-actions: \emph{publish} and \emph{subscribe}, which may involve uncertainties (e.g., the \emph{publish} and \emph{subscribe} actions may fail to execute due to port malfunctions) (see Section~\ref{subsec:uncertainty_profile} for more details).
Of course, \guillemotleft BeliefStatement\guillemotright\ can be directly applied to action \emph{publish} and \emph{subscribe} by restricting its scope.

\subsubsection{Indeterminacy Source}\label{subsubsec:indeterminacysource}

In PSUM, an indeterminacy source aims to describe a situation where the information required to be certain about the validity of a belief statement is indeterminate~\cite{omg2024psum}.
In \profile, we define stereotype \guillemotleft IndeterminacySource\guillemotright\ and allow it to be applied on occurrences and attributes. In SysML v2, an occurrence is a fundamental concept used to describe a class of anything that unfolds over time and may also have a spatial extent~\cite{omg2025sysmlv2}, such as a vehicle moving along a road or a person walking through a secured area. Many commonly used SysML v2 modeling elements, including parts, ports, states, and actions, are ultimately based on or derived from \textit{SysML::Occurrence}. In contrast, attributes are used to describe static properties of system elements rather than phenomena that occur or persist over time. Accordingly, \guillemotleft IndeterminacySource\guillemotright\ is defined to extend both occurrences and attributes in SysML v2.



\begin{definition}
    \textbf{\guillemotleft IndeterminacySource\guillemotright}
    \guillemotleft IndeterminacySource\guillemotright\ can be applied to \textit{SysML::OccurrenceDefinition, SysML::OccurrenceUsage, SysML::AttributeDefinition, SysML::AttributeUsage}, representing stereotyped elements as indeterminacy sources of uncertainties specified in belief statements.
\end{definition}

\textbf{(a) Part.}
A typical scenario for specifying indeterminacy sources is to stereotype components of a system whose correct functioning or operation is indeterminate, which may lead to uncertain system behavior.
As introduced above, the components of the system can be modeled as \textit{SysML::PartDefinition} and \textit{SysML::PartUsage}. Both specialize \textit{SysML::OccurrenceDefinition} and \textit{SysML::OccurrenceUsage}, respectively~\cite{omg2025sysmlv2}.
When we look at Example~\ref{example:acc}, multiple perception sensors are included in \emph{ACC}: \emph{radars}, \emph{lidars} and \emph{cameras}, which are labelled as indeterminacy sources (Lines~\ref{line:part_radar}-\ref{line:part_camera}), indicating that their outputs (e.g., object detection, range estimation, or image interpretation) are inherently imperfect and indeterminate, which form the sources of uncertainties in the behavior of the system.

Indeterminacy source can be further characterized by \textit{PSUM::IndeterminacyNature}, an enumeration with five literals: \textit{InsufficientResolution}, \textit{MissingInfo}, \textit{Non-determinism}, \textit{Unclassified} and \textit{Custom}~\cite{omg2024psum}.
For instance, indeterminacy sources \emph{cameras} and \emph{lidars} (which are part usages) are labelled as \textit{InsufficientResolution} (denoted by \textit{isr} in the textual notation (Lines~\ref{line:part_laser}-\ref{line:part_camera}) since a camera's limited spatial resolution may fail to resolve small objects and a lidar's point cloud sparsity under adverse weather may degrade perception accuracy. The indeterminacy of a sensor can also be characterized as non-determinism (denoted by \texttt{nd} in the textual notation, Line~\ref{line:part_radar}), as exemplified by part usage \emph{radars}. In this case, the sensor's output may be entirely absent due to unexpected physical obstruction.
Note that the choice of applying the stereotype to a definition or to a usage is, again, a methodological concern. In this example, we apply the stereotype to part usages rather than to part definitions, indicating that indeterminacy of radars, lidars, and cameras is only relevant in the specific context of their use within the adaptive cruise control system, rather than as a general property of these components in all contexts.


\textbf{(b) Port.}
In a system, indeterminacy may arise not only from the functioning of components modeled as parts, but also from their interactions, which can be explicitly manifested at ports and therefore treated as indeterminacy sources.
In SysML v2, \textit{SysML::PortDefinition} (or \textit{SysML::PortUsage}) specializes \textit{SysML::OccurrenceDefinition} (or \textit{SysML::OccurrenceUsage}). Signals exchanged through ports often control the execution of system actions or the triggering of state transitions. For example, a state transition may be triggered by a signal received through a specific port; indeterminacy in the port’s behavior or in signal transmission can therefore propagate into the system’s behavioral dynamics.

As depicted in Example~\ref{example:is}, port definitions \emph{PublicationPort} and \emph{SubscriptionPort} are labelled as the non-determinism kind of indeterminacy sources with \guillemotleft IndeterminacySource<nd>\guillemotright\ (Lines~\ref{line:port_def_pub} and~\ref{line:port_def_sub}), reflecting the fact that communication through these ports may be subject to unpredictable behavior, such as message loss, delays, or ports malfunction.
In this case, the indeterminacy sources are specified at the definition level and can be reused in different usage contexts, since, in an interaction sequence, the indeterminacy of ports in the context of each participating role needs to be considered.
As we discussed in Section~\ref{subsec:defUsagePair}, when a stereotype is applied to a definition element, it does not need to be redundantly applied to its usages; instead, all usages of that the definition element inherit the stereotype via the typing relationship. Specifically, port usages declared in part \emph{producer}, \emph{server}, and \emph{consumer} are also indeterminate because they are typed by their corresponding port definitions which are indeterminate.

The presence of these indeterminacy sources directly affects the system's behavior. For the \emph{producer}, indeterminacy in the \emph{publicationPort} introduces uncertainty in whether a \emph{publish} action successfully transmits publications to the \emph{server} (Line~\ref{line:act_publish}). Similarly, on the \emph{server} side, indeterminacy in both \emph{publicationPort} and \emph{subscriptionPort} affects the reliable reception of subscriptions and delivery of publications (Lines~\ref{line:trans_subscribing} and~\ref{line:trans_delivering}). For the \emph{consumer}, indeterminate \emph{subscriptionPort} leads to uncertainty in action for subscribing (Line~\ref{line:act_subscribe}). Overall, indeterminacy in port functionality could result uncertainties in action execution and interaction sequencing across the system.

\textbf{(c) Attribute.}
An attribute may be identified as an indeterminacy source when its value, type, multiplicity, or even its existence is incomplete, variable, or otherwise indeterminate. In SysML v2, an attribute specifies a set of data values, including numeric values, quantitative values with units, qualitative values like text strings, or data structures built from these values~\cite{omg2025sysmlv2}. Attributes can be specified with \textit{SysML::AttributeDefinition} or \textit{SysML::AttributeUsage}. 
\begin{example}
    For example, in the following state machine representing health state of a vehicle, the system transits to state \emph{maintenance} triggered by an absolute time \emph{maintenanceTime}. Thus, the attribute usage \emph{maintenanceTime} can be modeled as missing info kind of indeterminate source with «IndeterminacySource<mi>» (Lines~\ref{line:attr_maintenance_time}), indicating the full set of information about the maintenance time is unavailable when the belief statement is made.
    \begin{lstlisting}
part def Vehicle {
	«IndeterminacySource<mi>» attribute maintenanceTime: Time::DateTime; (*@\label{line:attr_maintenance_time}@*)
}
«BeliefStatement» state `health states' {
	part vehicle : Vehicle;
	entry action initial;
	transition initial then normal;
	state normal;
	transition `normal-maintenance'
	first normal accept at vehicle.maintenanceTime then maintenance;
	state maintenance;
}
    \end{lstlisting}
\end{example}

\subsubsection{Indetermiancy Specification}\label{subsubsec:indeterminacyspecification}

In \profile, we adopt the notion of \textit{IndeterminacySpecification} from U-Model proposed in~\cite{zhang2019uncertainty} to capture concrete functional or operational statuses of an indeterminacy source. \textit{PSUM::IndeterminacySource} is a subclass of \textit{PSUM::Belief}, and \textit{PSUM::Belief} has an composition relationship with \textit{PSUM::BeliefStatement}; hence \textit{U-Model::IndeterminacySpecification} is actually indirectly modeled as \textit{PSUM::BeliefStatement}.
To capture which alternative of an indeterminacy source occurs and to establish a formal linkage to uncertainty, our design introduces \guillemetleft IndeterminacySpecification\guillemetright\ explicitly.
By extending \textit{SysML::ConstraintUsage}, it enables indeterminacy alternatives to be precisely defined and evaluated as executable constraints within the model.

%

\begin{definition}
    \textbf{\guillemetleft IndeterminacySpecification\guillemetright}
    The stereotype can be applied to \textit{SysML::ConstraintUsage} owned by an indeterminacy source, to precisely define alternative specifications of the indeterminacy source, which can be evaluated as executable constraints.
\end{definition}

Example~\ref{example:is} contains two indeterminate port definitions: \emph{PublicationPort} and \emph{SubscriptionPort}. Each port has attribute \textit{operationalStatus} representing whether the port is functioning properly (Lines~\ref{line:attr_pub_status} and~\ref{line:attr_sub_status}). To distinguish the possible functional statuses, two specifications are defined in each port, modeled as \textit{SysML::ConstraintUsage} stereotyped with \guillemetleft IndeterminacySpecification\guillemetright\ (Lines~\ref{line:spec_pub_op}-\ref{line:map_pub_nop} and Lines~\ref{line:spec_sub_op}-\ref{line:map_sub_nop}). The \emph{Operational} constraint assumes that the port is functioning correctly (status == true), while the \emph{NotOperational} constraint represents the port being unavailable or failing (status == false).
By defining concrete indeterminacy specifications, uncertain behaviors can be traced to specific operational statuses of the underlying sources. For instance, the successful execution of the \emph{publish} action, which transmits publications to the \emph{server}, is conditioned on the \emph{publicationPortOperational} constraint (Line~\ref{line:ref_pub_port_op}), allowing the model to explicitly capture the impact of different statuses in indeterminacy sources on system behaviors.

\subsection{The Uncertainty Profile}\label{subsec:uncertainty_profile}
\begin{figure}[H]
    \centering
    \includegraphics[width=0.9\textwidth]{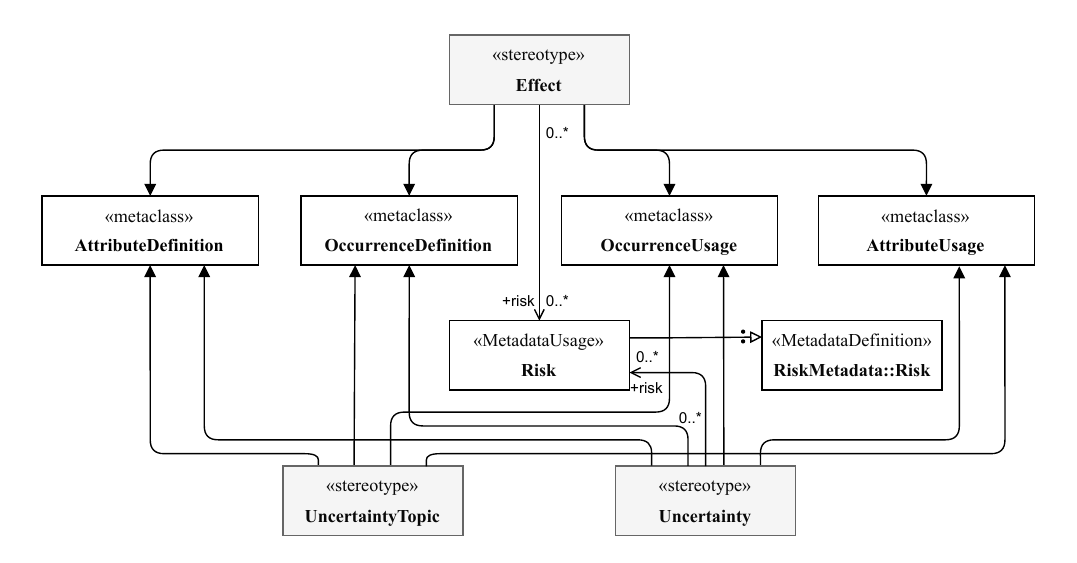}
    \caption{The \profile uncertainty profile. The classes in grey are stereotypes while those in white are SysML v2 metaclasses}\label{fig:uncertainty_profile}
\end{figure}

The uncertainty profile aims to implement the PSUM uncertainty metamodel package. Especially, an uncertainty represents a deficiency of information or knowledge needed to assess the content, consequences, or likelihood of an uncertainty topic existed in a belief statement~\cite{omg2024psum}.
Note that we do not need to explicitly capture \textit{PSUM::UncertaintyHistory}, which is used for keeping the record of a sequence of changes applied to uncertainty characteristic or measurement of an uncertainty~\cite{omg2024psum}. Instead, the management of uncertainty history can be delegated to the supporting tool infrastructure, for instance through database storage.
%
Therefore, as shown in Figure~\ref{fig:uncertainty_profile}, we define three stereotypes: \guillemotleft Uncertainty\guillemotright, \guillemotleft UncertaintyTopic\guillemotright\ and \guillemotleft Effect\guillemotright.
\textit{Risk} has been implemented in the standard metadata domain library of SysML v2; hence we can directly leverage it without introducing new notations.

\subsubsection{Uncertainty}\label{subsubsec:uncertainty}

Uncertainty is one of the key concepts in PSUM and hence \profile. We define stereotype \guillemotleft Uncertainty\guillemotright\ that can be applied to \textit{SysML::Occurrence} and \textit{SysML::Attribute}.
The rationale is that: together, \textit{SysML::Occurrence} and \textit{SysML::Attribute} correspond to two fundamental modeling dimensions: temporally and spatially extended existence (what happens or exists, over time and potentially across space) and descriptive characterization (what specifies, quantifies, or constrains it).
In SysML v2, an occurrence represents individuals that persist over time as identifiable entities and may also possess spatial extent.
Consequently, uncertainty related to the execution of actions, the firing of transitions, lifecycle, temporal extent, and behavioral interaction is naturally anchored in occurrences, since these represent entities that unfold over time and potentially across space.
Complementarily, \textit{SysML::Attribute} captures the properties and parameters that characterize such occurrences, providing the extension point for expressing uncertainty in values, ranges, probabilistic specifications, etc. By extending only these two metaclasses, the mechanism remains minimal yet semantically comprehensive, avoiding fragmentation while aligning directly with the core abstractions of SysML v2 semantics.

Below, we discuss the application of the stereotype to transitions, actions, and attributes, as these elements are particularly representative and illustrative, especially in supporting instance-level analysis~\cite{burgueno2023dealing} and model-based testing scenarios~\cite{zhang2019uncertainty}.

\begin{definition}
    \textbf{\guillemotleft Uncertainty\guillemotright}
    The stereotype can be applied to \textit{SysML::OccurrenceDefinition, SysML::OccurrenceUsage, SysML::AttributeDefinition, SysML::AttributeUsage} to denote that stereotyped elements are uncertain. Such uncertain elements can be further characterized by uncertainty perspectives, patterns, or effects, and associated with indeterminacy sources and their specifications.
\end{definition}

\textbf{(a) Transition.}
When a state machine is declared as belief statement, its transitions could be stereotyped with \guillemotleft Uncertainty\guillemotright, indicating that the modeler does not have full confidence that the source state will transit to the target state via the corresponding transition, given the guard condition and trigger (if any).

As shown in Example~\ref{example:acc}, after belief statements (Section~\ref{subsubsec:beliefstatement}) and indeterminacy sources (Section~\ref{subsubsec:indeterminacysource}) are captured, we can specify uncertain elements of the belief statements and associate them to the corresponding indeterminacy sources.
In the belief state machine \emph{decisionLayerState}, transition \emph{startDeciding} from state \emph{waitingForSignal} to state \emph{deciding}, which is triggered upon receiving  \emph{PerceptionSignal}, is stereotyped with \guillemotleft Uncertainty\guillemotright\ (Line~\ref{line:trans_start_deciding}). This indicates that the transition is uncertain due to the indeterminate outputs of perception sensors (e.g., \emph{radars}) defined in the part definition (Line~\ref{line:part_radar}). Specifically, the indeterminacy specification of this transition is referred to the \emph{radarNotBlocked} constraint of the indeterminacy source \emph{radar} (Line~\ref{line:ref_not_blocked}), representing that \emph{PerceptionSignal} can trigger the transition normally only if the radar is not blocked by unexpected physical obstructions.
%
Uncertain transition \emph{failToStartDeciding} (Line~\ref{line:trans_fail_deciding}) is defined to capture the situation that the radar is blocked. Under this condition, \emph{PerceptionSignal} cannot trigger the transition from state \emph{waitingForSignal} to \emph{deciding}; instead, the perception layer remains in state \emph{waitingForSignal}. In contrast to \emph{startDeciding}, the indeterminacy specification of this transition is referred to constraint \emph{radarBlocked} (Line~\ref{line:ref_blocked}).

An uncertainty can be characterized by uncertainty kind, nature and reducibility.
%
%
\textit{PSUM::UncertaintyKind} is defined to categorize different sources of uncertainty from the perspective of a belief agent. It includes: \textit{ContentUncertainty}, which arises when the agent is unsure about the factual content of a belief statement; \textit{EnvironmentUncertainty}, reflecting doubt regarding the physical surroundings described in the statement; \textit{GeographicalLocationUncertainty}, indicating ambiguity about a location referenced in the statement; \textit{OccurrenceUncertainty}, capturing uncertainty about whether an event existed in belief statement occurs; and \textit{TimeUncertainty}, denoting imprecision or lack of confidence concerning temporal information within the statement. In Example~\ref{example:acc}, transitions \emph{startDeciding} and \emph{failedToStartDeciding} contain occurrence uncertainties (textually denoted by \textit{ocr}, in Lines~\ref{line:trans_start_deciding} and~\ref{line:trans_fail_deciding}), indicating that the uncertainties lie in whether the corresponding transition actually occurs or not, which depends on the physical status of \emph{radars}.

\textit{PSUM::UncertaintyNature} specifies whether an uncertainty is aleatory or epistemic. Aleatory uncertainty stems from the intrinsic randomness or natural variability of real-world phenomena, which cannot be reduced even with additional knowledge. In contrast, epistemic uncertainty arises from incomplete or imperfect information~\cite{omg2024psum}. In Example~\ref{example:acc}, \emph{startDeciding} and \emph{failedToStartDeciding} are classified as epistemic uncertainty within \guillemotleft Uncertainty\guillemotright\ (textually denoted by \textit{epi}, in Lines~\ref{line:trans_start_deciding} and~\ref{line:trans_fail_deciding}), since the uncertainties originate from incomplete knowledge about the operational condition of the perception sensor \emph{radar} rather than from inherent randomness of the environment. Such uncertainties could be reduced or resolved by acquiring more accurate or reliable sensing information, for example through redundant sensors.

For epistemic uncertainty, it could be further categorized by \textit{PSUM::ReducibilityLevel} based on its potential for reduction: \textit{FullyReducible} uncertainty denotes situations where full certainty is initially absent but can be achieved by collecting additional information; \textit{PartiallyReducible} uncertainty refers to cases where full certainty is absent but uncertainty can still be reduced through additional information; and \textit{Irreducible} uncertainty describes situations in which full certainty is absent and cannot be reduced~\cite{omg2024psum}. Based on the defined reducibility level, as shown in Example~\ref{example:acc}, \emph{startDeciding} and \emph{failedToStartDeciding} are classified as partially reducible uncertainty (Lines~\ref{line:u_reducibility_1} and~\ref{line:u_reducibility_2}), meaning that they can be partially mitigated by collecting additional information through sensor redundancy. However, it is not fully reducible, since even redundant sensors could be simultaneously blocked by unexpected physical obstructions, meaning that full certainty cannot be guaranteed in all possible scenarios.

In addition to uncertainty kind, nature, and reducibility, each uncertainty must be characterized as either \textit{subjective} or \textit{objective}. The subjective uncertainty perspective refers to information known or inferred by an observing agent through observation or reasoning, whereas the objective uncertainty perspective concerns phenomena or concepts whose existence and nature are independent of any observing agent~\cite{omg2024psum}.
In Example~\ref{example:acc}, as the occurrence of uncertain transitions \emph{startDeciding} and \emph{failedToStartDeciding}
are not supported by concrete evidence and is specified based on the expertise and experience of the belief agent, they are  characterized as subjective uncertainty (textually denoted by \textit{subj},
in Lines~\ref{line:trans_start_deciding} and~\ref{line:trans_fail_deciding}).
However, this characterization can evolve over time, e.g., shifting from subjective to objective once all supporting evidence becomes available.

And lastly, for \textit{PSUM::OccurrenceUncertainty}, a pattern could be defined. The possible pattern types, as defined in PSUM~\cite{omg2024psum}, include \textit{periodic}, \textit{persistent}, \textit{sporadic}, \textit{transient} and \textit{random}. Therefore, as shown in Example~\ref{example:acc}, \emph{startDeciding} and \emph{failedToStartDeciding} are assigned the random pattern (Lines~\ref{line:u_pattern_1} and~\ref{line:u_pattern_2}), as the occurrence of physical obstructions affecting \emph{radars} is inherently unpredictable and driven by stochastic environmental conditions.

\textbf{(b) Action.}
As discussed in Section~\ref{subsubsec:beliefstatement}, when an action is declared as a belief statement, its parameters and sub-actions (e.g., send action, accept action, assignment action) may be specified as uncertain, reflecting potential uncertainties in actual performance of the action.
For instance, as illustrated in Example~\ref{example:is}, action \emph{producerBehavior} (Line~\ref{line:act_prod_behavior}), which has been stereotyped as belief statement (Section~\ref{subsubsec:beliefstatement}), a send action inside is responsible for sending publications of certain topics from \emph{producer} to \emph{server} via \emph{publicationPort} (Line~\ref{line:port_prod_pub}), which has been defined by an indeterminacy source (Section~\ref{subsubsec:indeterminacysource}). Send action \emph{publish} (Line~\ref{line:act_publish}) is therefore stereotyped with \guillemotleft Uncertainty\guillemotright, reflecting that its execution may be uncertain due to the indeterminate operational status of \emph{publicationPort}. The associated indeterminacy specification is referred to constraint \emph{publicationPortOperational} (Line~\ref{line:ref_pub_port_op}), indicating that successful transmission depends on the correct functioning of the port. For \emph{consumer}, its action \emph{subscribe} in belief statement \emph{consumerBehavior}, responsible for subscribing specific topics to the \emph{server}, is modeled as an uncertain action (Line~\ref{line:act_subscribe}), with indeterminacy specification refers to constraint \emph{subscriptionPortOperational} (Line~\ref{line:ref_cons_sub_op}).
Similarly, the uncertain actions are further characterized as occurrence, epistemic and subjective uncertainty (Lines~\ref{line:act_publish} and~\ref{line:act_subscribe}).

\textbf{(c) Attribute.}
As illustrated in Example~\ref{exampe:vfea}, attribute usage \emph{wheelDiameter} of \emph{Vehicle}, typed by \textit{SysML::ISQ::LengthValue}, cannot be represented as a fixed concrete value, as it varies due to manufacturing tolerances, wear, and operating conditions, which is an importance concern of this modeling activity.
Hence, this attribute is stereotyped as uncertain (Line~\ref{line:attr_wheel_dia}).
Its uncertainty kind is specified as content uncertainty (textually denoted by \textit{con}, in Line~\ref{line:attr_wheel_dia}), with an aleatory nature (textually denoted by \textit{ale}, in Line~\ref{line:attr_wheel_dia}).
To precisely characterize this content uncertainty, a \textit{measurement} block is defined (Lines~\ref{line:meas_start}–\ref{line:meas_end}), specifying a measurement tolerance of approximately 1.5\% relative to the assumed nominal wheel diameter value (Line~\ref{line:meas_error}).
This measurement error represents the expected deviation between the modeled value and the actual physical value of the wheel diameter.
This \textit{uncertain} attribute is further employed in constraint \emph{fuelEconomyAnalysisAssumedConstraint} declared as an assumed constraint (Line~\ref{line:const_assume}), which defines the assumed conditions under which the fuel economy analysis is conducted.
Within this constraint, attribute \emph{vehicle.wheelDiameter}, originally assumed to have a nominal value of \textit{33 [inch]} (Line~\ref{line:expr_dia_original}), is instead represented as a bounded range between \textit{32.505 [inch]} and \textit{33.495 [inch]} (Lines~\ref{line:expr_dia_min}–\ref{line:expr_dia_max}) based on the specified  tolerance around the nominal value.
This update explicitly captures the uncertainty defined by the attribute stereotype, enabling further systematic uncertainty-aware analysis and evaluation.

\subsubsection{UncertaintyTopic}\label{subsubsec:uncertainty_topic}

\textit{PSUM::UncertaintyTopic} is for relating an uncertainty to associated belief statements~\cite{omg2024psum}; however, in \profile, we treat uncertainty topic solely as a user-defined construct for explicitly organizing uncertainties. Since uncertainty has been directly specified within model elements stereotyped as belief statements, the association between uncertainty and belief statement does not need to be established through uncertainty topic in this context. The issue of whether the same uncertainty topic can be discussed in multiple belief statements described using different languages is out of the scope of this work, as we only focus on modeling with SysML v2.

\begin{definition}
    \textbf{\guillemotleft UncertaintyTopic\guillemotright}
    The stereotype can be applied to \textit{SysML::OccurrenceDefinition}, \textit{SysML::OccurrenceUsage}, \textit{SysML::AttributeDefinition}, \textit{SysML::AttributeUsage} to declare a topic of interest to the modeler, which may be associated to related uncertainties.

\end{definition}

For instance, in Example~\ref{example:acc}, we declare an uncertainty topic of item definition \emph{PerceptionSignal} (Line~\ref{line:def_percep_signal}), with related uncertainties refer to uncertain transition \emph{startDeciding} and \emph{failToStartDeciding} (Lines~\ref{line:ref_u_start}-\ref{line:ref_u_fail}). The topic captures uncertain transitions that are triggered by \emph{PerceptionSignal}.
Similarly, in Example~\ref{example:is}, we apply \guillemotleft UncertaintyTopic\guillemotright\ to item definitions \emph{Subscribe}, \emph{Publish} and \emph{Deliver} (Lines~\ref{line:item_subscribe},~\ref{line:item_publish},~\ref{line:item_deliver}), each grouping uncertain transitions that are triggered by the signal or uncertain send actions taking the signal as payload (Lines~\ref{line:u_ref_sub_server}-\ref{line:u_ref_sub_consumer},~\ref{line:u_ref_pub_producer}-\ref{line:u_ref_pub_server},~\ref{line:u_ref_del_server}-\ref{line:u_ref_del_consumer}).

\subsubsection{Effect}\label{subsubsec:effect}

\textit{PSUM::Effect} is a special kind of \textit{PSUM::UncertaintyCharacteristic} used to capture the result of certain uncertainty. In \profile, we design stereotype \guillemotleft Effect\guillemotright\ for it, which extends the same SysML v2 metaclasses as \guillemotleft Uncertainty\guillemotright, since \textit{PSUM::Effect} is defined as a subtype of \textit{PSUM::Uncertainty}~\cite{omg2024psum}. Capturing the association between uncertainty and its associated effect can be useful for specifying the propagation of uncertainty.

\begin{definition}
    \textbf{\guillemotleft Effect\guillemotright}
    This stereotype can be applied to \textit{SysML::OccurrenceDefinition}, \textit{SysML::OccurrenceUsage}, \textit{SysML::AttributeDefinition}, \textit{SysML::AttributeUsage} to denote that the stereotyped element is a result of a specified uncertainty.

\end{definition}

As illustrated in Example~\ref{example:is}, the uncertainty of send action \emph{publish} (Line~\ref{line:act_publish}) propagates to transition \emph{delivering} in state \emph{serverBehavior} through the interaction between \emph{producer} and \emph{server}. Specifically, \emph{producer} sends publications to \emph{server}, triggering a transition that accepts the publications and delivers them to \emph{consumer}. As a result, transition \emph{delivering} can be specified as an effect of uncertain action \emph{publish} (Lines~\ref{line:ref_eff_del} and~\ref{line:trans_delivering}). Furthermore, upon receiving publications from \emph{server}, action \emph{delivery} of \emph{consumer} is triggered, which is also stereotyped as an effect of \emph{delivering} (Lines~\ref{line:ref_eff_cons_del} and~\ref{line:act_delivery}). This interaction sequence models a two-level propagation of uncertainty: from action \emph{publish}, to transition \emph{delivering}, and subsequently to action \emph{delivery}. Similarly, the action \emph{subscribe} (Line~\ref{line:act_subscribe}) of \emph{consumerBehavior}, which is responsible for subscribing to a specific topic on \emph{server}, triggers the transition \emph{subscribing} from the state \emph{waitForSubscription} to \emph{waitForPublication}. Accordingly, \guillemotleft Effect\guillemotright\ is applied to \emph{subscribing} to indicate that it is the effect of action \emph{subscribe} (Line~\ref{line:trans_subscribing}).

Through the association, the indeterminacy sources and specifications of an uncertainty can also propagate to its effect. For example, as shown in Lines~\ref{line:ref_prod_port_op}-\ref{line:ref_srv_pub_op2}, the indeterminacy sources and specifications of \emph{delivering} are defined by constraint \emph{publicationPortOperational} for both \emph{server} and \emph{producer}, indicating that the successful execution of the transition depends not only on the correct functioning of \emph{server}'s \emph{publicationPort}, but also on that of \emph{producer}. We explicitly model \emph{producer.publicationPort.publicationPortOperational} here for illustration (Line~\ref{line:ref_prod_port_op}). In practice, however, the specification should be derived from the association between an uncertainty and its effect.


\subsubsection{Risk}\label{subsubsec:risk}

As defined In PSUM, when an uncertainty is introduced into the system, it inherently introduces \textit{PSUM::Risk} which is defined as the effect of uncertainty on objectives~\cite{ISO31000}.
When specifying Risk within \profile, we do not introduce it as an individual stereotype. Instead, we directly leverages the risk modeling facilities provided by the standard model library of SysML v2. Specifically, the standard library package \emph{SysML::RiskMetadata} defines a metadata definition \emph{SysML::Risk}, which is intended to annotate model elements with assessments of risk, including overall risk levels as well as optional technical, schedule, and cost risks~\cite{omg2025sysmlv2}. Therefore, the risk associated with an uncertain model element can be directly represented by annotating it with a metadata usage defined by \emph{RiskMetadata::Risk}, rather than defining a custom risk stereotype. This approach avoids semantic duplication and ensures consistency with the standardized risk concepts and levels provided by SysML v2.

\begin{definition}
    \textbf{Risk}
    The risk of an uncertain model element is represented as \textit{SysML::MetadataUsage} defined by \emph{RiskMetadata::Risk}, where the uncertain element is specified as the annotated element of this metadata usage.

\end{definition}

For instance, in Example~\ref{example:acc}, we specify \emph{collisionRisk} (Line~\ref{line:risk_collision}) for uncertain transition \emph{failToStartDeciding} in state \emph{decisionLayerState}. This risk indicates that, when the radar is blocked, the decision layer may fail to receive perception signals required to trigger the transition for computing control commands (e.g., a deceleration command), which may in turn lead to a potential collision with leading vehicle. The associated impact level of \emph{collisionRisk} is set to \emph{high} (Lines~\ref{line:risk_impact}), as defined in \emph{RiskMetadata::LevelEnum}, reflecting the potentially severe consequences of a collision.

\subsection{The Measurement Profile}\label{subsec:measurement_profile}

The measurement profile of \profile aims to implement the PSUM measurement metamodel package to provide a generic approach to quantify measurable elements of the PSUM metamodel.
\textit{PSUM::MeasurableElement} refers to \textit{PSUM::PSUMElement} that can be measured. In this work, we concern about the measurement of \textit{PSUM::BeliefStatement}, \textit{PSUM::IndeterminacySource} and \textit{PSUM::Uncertainty}. \textit{PSUM::Risk} is realized through \emph{SysML::RiskMetadata::Risk} and can be measured in standard level provided in the library (Section~\ref{subsubsec:risk}).
For each measurable element, it may be composed of a set of associated measurable features, i.e., multiple aspects may be quantified to characterize an element, including \textit{PSUM::Accuracy}, \textit{PSUM::Sensitivity}, \textit{PSUM::MeasurementError}, \textit{PSUM::Precision}, and \textit{PSUM::Degree}~\cite{omg2024psum}.
In \profile, the measurable features are represented as classes that are instantiated when specified in a measurable element, i.e., a SysML model element to which a measurable stereotype has been applied.

According to PSUM, a measurable feature should be associated with a measurement from the Structured Metrics Metamodel (SMM) specification~\cite{omg2025smm}. However, in the context of SysML v2, it provides a standard model library, Quantities and Units Domain Library, which defines reusable and extensible model elements for physical quantities, including vector and tensor quantities, quantity dimensions, measurement units, measurement scales, coordinate frames, coordinate transformations and vector spaces. The library also enables the specification of coherent systems of quantities and systems of units, as well as the operators and functions needed to support quantity arithmetic in expressions~\cite{omg2025sysmlv2}. Therefore, in \profile, a \textit{PSUM::MeasurableFeature}'s measurement is designed to be \textit{KerML::Expression},
which can utilize the standard library of SysML v2 to formally measure different measurable features of SysML model elements with measurable stereotype. Utilizing the native modeling capability of SysML v2 facilitates efficient measurement of measurable features, and ensures that the representation of measurement remains consistent with the way feature values are expressed in SysML models.

\begin{definition}
    \textbf{Measurement}
    The measurement of an measurable feature is represented as \textit{KerML::Expression}, quantifying different aspects of SysML v2 model elements stereotyped with \guillemotleft BeliefStatement\guillemotright, \guillemotleft IndeterminacySource\guillemotright\ or \guillemotleft Uncertainty\guillemotright.
\end{definition}

As shown in Example~\ref{exampe:vfea}, where attribute \emph{wheelDiameter} with content uncertainty has been introduced (Section~\ref{subsubsec:uncertainty}), we can further specify the error range of \emph{wheelDiameter} caused by inherent real-world randomness as 1.5\% of the assumed diameter value (\emph{0.495[inch]} in this case) in measurement error (Line~\ref{line:meas_error}). Subsequently, we update constraint \emph{fuelEconomyAnalysisAssumedConstraint}, which originally asserted a precise value of \emph{33[inch]} (Line~\ref{line:expr_dia_original}), to allow for a range from \emph{32.505[inch]} to \emph{33.495[inch]} (Lines~\ref{line:expr_dia_min}-\ref{line:expr_dia_max}).

\section{Case Studies}\label{sec:case_study}

This section introduces the selection criteria for the case studies, presents the results analysis, and discusses the benefits, limitations, and threats to validity in assessing \profile.

\subsection{Case Study Selection}\label{subsec:case_description}

To evaluate \profile, we selected \sysml models from publicly available repositories. 
Specifically, we checked all \sysml examples provided by the SysML v2 working group , i.e., the SysMLv2 Release repository~\cite{SysMLv2-Release}. 
In addition, we searched for additional available \sysml models online, including those in the SysML-v2-Models repository~\cite{GfSE_SysMLv2_Models_2025}.
\begin{table}
	\centering
			\centering
			\caption{Overall descriptive statistics of the case studies}
			\label{tab:case_studies}
			\begin{tabular}{lrrrrrrrr}
				\toprule
				Case study & ACC & IS & VFEA & AF & VM & DS & MF & \textbf{\#Total} \\
				\midrule
				Lines of Models (LoM) & 190 & 144 & 286 & 332 & 1579 & 407 & 747 & \textbf{3685} \\\hline
				\#requirement def       & -   & -   & 1   & -   & 5    & 10   & -   & \textbf{16}    \\
				\#requirement       & -   & -   & 8   & -   & 14    & 13   & -   & \textbf{35}    \\
				\#analysis def       & -   & -   & 1   & -   & -    & -   & -   & \textbf{1}    \\
				\#analysis       & -   & -   & 2   & -   & 2    & -   & -   & \textbf{4}    \\
				\#part def             & 9   & 1   & 3   & 6   & 42   & 10  & 26  & \textbf{97}   \\
				\#part                 & 8   & 11  & 5   & 6   & 110  & 34  & 27  & \textbf{201}  \\
				\#item def             & 6   & 3   & 1   & -   & 9    & -   & 12  & \textbf{31}   \\
				\#item                 & -   & 1   & 3   & -   & 19   & 4   & 8   & \textbf{35}   \\
				\#port def             & -   & 2   & 1   & 8   & 27   & -   & 5   & \textbf{43}   \\
				\#port                 & -   & 4   & 5   & 19  & 86   & 2   & 10  & \textbf{126}  \\
				\#attribute def        & -   & -   & 5   & 7   & 9    & 9   & -   & \textbf{30}   \\
				\#attribute            & 17  & 4   & 25  & 16  & 123  & 11  & 61  & \textbf{257}  \\
				\#action def           & -   & -   & 1   & -   & 8    & -   & 17  & \textbf{26}   \\
				\#action               & 5   & 5   & 2   & 9   & 54   & 3   & 17  & \textbf{95}   \\
				\#state def            & 1   & -   & -   & -   & 23   & 2   & -   & \textbf{26}   \\
				\#state                & 15  & 3   & -   & 10  & 25   & 4   & 6   & \textbf{63}   \\
				\#transition           & 11  & 4   & -   & 10  & 15   & 8   & 17  & \textbf{65}   \\
				\#message              & -   & 6   & -   & -   & 6    & 1   & 23  & \textbf{36}   \\
				\#occurrence def       & -   & -   & -   & -   & -    & -   & 4   & \textbf{4}    \\
				\bottomrule
			\end{tabular}
			\begin{flushleft}
				\footnotesize
				\textbf{Note:} The statistics are collected based on keyword occurrences. Keywords containing \emph{def} denote definition elements, whereas keywords without \emph{def} (e.g., requirement, part, action, state) denote their corresponding usage elements.
			\end{flushleft}
\end{table}
Given our research context, we excluded toy examples, and included those with both structural and behavioral aspects.
This process resulted in seven case studies:
(1) Adaptive Cruise Control (ACC),
(2) Interaction Sequencing (IS),
(3) Vehicle Fuel Economy Analysis (VFEA),
(4) Arrowhead Framework (AF),
(5) Vehicle Model (VM),
(6) Drone System (DS), and
(7) Mining Frigate (MF).
ACC, IS, and VFEA served as the running examples (Section~\ref{sec:running examples}) throughout this paper and are therefore not described further.

The AF~\cite{SysMLv2-Release} model represents a hybrid communication architecture combining publish–subscribe messaging and asynchronous remote procedure calls, based on an industrial Arrowhead Framework deployment for a chemical factory under the Productive 4.0 initiative.

The VM~\cite{SysMLv2-Release} model provides a comprehensive vehicle system representation centered on the \emph{Vehicle} part. 
It integrates structural, behavioral, requirements, allocation, analysis, and verification aspects. 
Structurally, the vehicle is decomposed into subsystems such as the powertrain, fuel system, control software, sensing, and interior components, with logical elements allocated to physical realizations to enable multi-level traceability.

The DS~\cite{GfSE_SysMLv2_Models_2025} model describes a configurable drone system integrating mechanical, electrical, and software components. 
The abstract part \emph{Drone} aggregates subsystems including battery, engine, body, and flight control, along with associated control, safety, and autonomous flight modules.

The MF~\cite{GfSE_SysMLv2_Models_2025} model represents a virtual mining frigate system from an online game.
The model includes a fixed \emph{CoreSystem} and configurable modules such as mining, defense, propulsion, sensing, and storage components. 
The frigate exposes ports for control, docking, defense, and ore handling, and references cargo and drone assets.

Descriptive statistics of the selected \sysml models are presented in Table~\ref{tab:case_studies}. 
Since we directly worked on the \sysml textual notations, we report the Lines of Models (LoM) as a measure of model size and an indicator of model complexity. As shown in Table~\ref{tab:case_studies}, LoM of the models ranges from 144 to 1579. 
In addition, we report the counts of core modeling elements representing requirement, structural and behavioral aspects, such as \emph{requirement def}, \emph{requirement}, \emph{part def}, \emph{part}, \emph{action def}, \emph{action}, \emph{state def}, and \emph{state}.
Overall, our case studies were conducted on seven \sysml models comprising a total of 3,685 LoM and 1,191 core modeling elements. 
These include 16 \emph{requirement def} and 35 \emph{requirement} elements, 97 \emph{part def} and 201 \emph{part} elements, 26 \emph{action def} and 95 \emph{action} elements, and 26 \emph{state def} and 63 \emph{state} elements, which reflects a broad coverage of requirement, structural, and behavioral modeling aspects across models of varying size and complexity.

\subsection{Results and Analyses}\label{subsec:results}


To assess \profile, we report statistics on core \sysml elements stereotyped using the profile. 
More specifically, Table~\ref{tab:belief_statement} presents the number of belief statements introduced in each case study, Table~\ref{tab:uncertainty} summarizes the number of uncertainty annotations and their associated effects, and Table~\ref{tab:indeterminacy_source} reports the number of indeterminacy sources and their corresponding specifications.

\begin{table}
	\centering
			\centering
			\caption{Number of belief statements introduced in each case study}
			\label{tab:belief_statement}
			\begin{tabular}{lrrrrrrr}
				\toprule
				\multirow{2}{*}{Case study} 
				& \multicolumn{5}{c}{\#Stereotyped element} 
				& \multirow{2}{*}{\#Total} \\
				
				\cmidrule(lr){2-6}
				
				& \#part def & \#part & \#action & \#state & \#occurrence def &  \\ 
				\midrule
				ACC  & -      & -      & 1 (25)  & 2 (223)  & -       & 3 (248)  \\
				IS   & -      & 1 (67) & 2 (45)  & 1 (99)   & -       & 4 (211)  \\
				VFEA & 1 (30) & 3 (19) & -       & -        & -       & 4 (49)   \\
				AF   & -      & -      & 1 (22)  & 4 (325)  & -       & 5 (347)  \\
				VM   & -      & 2 (90) & 1 (28)  & 3 (170)  & -       & 6 (288)  \\
				DS   & -      & -      & -       & 1 (224)  & -       & 1 (224)  \\
				MF   & -      & -      & -       & 1 (811)  & 4 (492) & 5 (1303) \\
				\midrule
				\textbf{Total} & \textbf{1 (30)} & \textbf{6 (176)} & \textbf{5 (120)} & \textbf{12 (1852)} & \textbf{4 (492)} & \textbf{28 (2670)} \\
				\bottomrule
			\end{tabular}
			\begin{flushleft}
				\footnotesize
				\textbf{Note:} Numbers in () are LoM of elements extended with \profile. 
			\end{flushleft}
\end{table}

\begin{table}
	\centering
			\centering
			\caption{Number of uncertainties and effects introduced in each case study}
			\label{tab:uncertainty}
			\begin{tabular}{lrrrrrrrr}
				\toprule
				\multirow{2}{*}{Case study} 
				& \multicolumn{4}{c}{\#Stereotyped element} 
				& \multirow{2}{*}{\#Total} 
				& \multirow{2}{*}{\#Effect} \\
				
				\cmidrule(lr){2-5}
				
				& \#attr  & \#action  & \#transition & \#message &  &  \\
				\midrule
				ACC    & -       & 1 (22)  & 6 (148)  & -        & 7 (170)   & 3  \\
				IS     & -       & 2 (26)  & 4 (88)   & 2 (36)   & 8 (150)   & 3  \\
				VFEA   & 1 (4)   & -       & -        & -        & 1 (4)     & -  \\
				AF     & -       & 4 (52)  & 11 (248) & -        & 15 (300)  & 7  \\
				VM     & -       & 2 (10)  & 7 (101)  & 4 (27)   & 13 (138)  & 1  \\
				DS     & -       & -       & 16 (203) & -        & 16 (203)  & -  \\
				MF     & -       & -       & 32 (780) & 23 (347) & 55 (1127) & -  \\
				\midrule
				\textbf{Total} & \textbf{1 (4)} & \textbf{9 (110)} & \textbf{76 (1568)} & \textbf{29 (410)} & \textbf{115 (2092)} & \textbf{14}\\
				\bottomrule
			\end{tabular}
			\vspace{5pt}
			\begin{flushleft}
				\footnotesize
				\textbf{Note:} 
				Numbers in () are LoM of elements extended with \profile; attr: attribute; LoM for \textit{Effect} are not collected, as all effects are further regarded as uncertainties (i.e., cause–effect chains) that have already been collected.
			\end{flushleft}
\end{table}

Overall, \profile was applied to a diverse set of structural and behavioral elements across the seven case studies. 
As shown in Table~\ref{tab:belief_statement}, a total of 28 elements were extended with «BeliefStatement», primarily targeting behavioral modeling elements, such as \emph{state}, \emph{action} and \emph{occurrence def} elements, covering 2,670 LoM.
Belief statements were stereotyped on occurrence definitions, which appear only in MF, for denoting enterprise-level interaction sequences such as \textit{DeployDrones} and \textit{OffloadOreAndResupply}. 
These occurrence definitions stereotyped as «BeliefStatement»  establish the interaction context in which individual messages are exchanged, allowing those messages to be explicitly annotated with «Uncertainty» to represent uncertainty characteristics and associated indeterminacy constraints (see Table~\ref{tab:uncertainty}).
In structural modeling, a part definition in VFEA is stereotyped with «BeliefStatement» to represent assumed structural properties of system components, such as part \textit{Vehicle}, including attributes like mass, drivetrain efficiency, and fuel economy, as discussed in Section~\ref{subsubsec:beliefstatement}. 
Within this part definition stereotyped as «BeliefStatement», specific attributes, such as \textit{wheelDiameter}, are further annotated with «Uncertainty» to explicitly capture content-related uncertainty arising from physical variability, along with associated measurement specifications.
The «BeliefStatement» stereotype applied to a part is also associated with its behavioral aspects. 
For example, in VM, part \textit{vehicle\_b } in the \textit{CruiseControl2} occurrence defines behavioral interactions through message exchanges such as \textit{sendSensedSpeed} and \textit{sendFuelCmd}.

Results of applying the uncertainty profile are summarized in Table~\ref{tab:uncertainty}. 
Overall, 115 elements, spanning 2,092 LoM, were annotated with «Uncertainty», most commonly applied to \emph{transition} and \emph{message} elements, with 14 associated effects representing the potential propagation and consequences of uncertainty.
Across the case studies, MF contributes 55 of the 115 uncertain elements, primarily associated with transitions and messages, demonstrating a broad operational treatment of uncertainty in interaction-driven and mission-level behaviors.
DS, AF, and VM provide a moderate number of uncertainty annotations.
More specifically, DS focuses on state transition uncertainty in drone operations, AF emphasizes uncertainty in system interactions and communication, and VM integrates uncertainty across states, actions, and messages within a large, mixed structural–behavioral model. 
ACC  and IS contribute smaller yet representative sets of uncertainty annotations, while VFEA shows that one focused analytical model can be extended with minimal uncertainty analysis.

\begin{table}
	\centering
	\resizebox{\textwidth}{!}{%
		\begin{minipage}{\textwidth}
			\centering
			\caption{Number of indeterminacy sources and specifications introduced in each case study}  
			\label{tab:indeterminacy_source}
			\resizebox{.99\linewidth}{!}{
			\begin{tabular}{lrrrrrrrrrrr}
				\toprule
				\multirow{2}{*}{Case study} 
				& \multicolumn{6}{c}{\#Stereotyped element} 
				& \multicolumn{3}{c}{\#Nature} 
				& \multirow{2}{*}{\#Total} 
				& \multirow{2}{*}{\#IndSpec} \\
				
				\cmidrule(lr){2-7}
				\cmidrule(lr){8-10} 
				
				& \#part def & \#part & \#port def & \#port & \#attr & \#occur def
				& \#ISR & \#MI & \#ND 
				&   &   \\
				
				\midrule
				ACC    & -      & 6 (40)  & -      & -      & -      & -      & 2 & - & 4  & 6 (40)    & 8   \\
				IS     & -      & -       & 2 (24) & 4 (4)  & -      & -      & - & - & 6  & 6 (28)    & 12  \\
				VFEA   & -      & -       & -      & -      & -      & -      & - & - & -  & -         & -   \\
				AF     & -      & -       & -      & 14 (29)& -      & -      & - & - & 14 & 14 (29)   & 28  \\
				VM     & -      & -       & 4 (50) & 7 (9)  & 2 (18) & -      & - & 2 & 11 & 13 (77)   & 26  \\
				DS     & -      & 8 (120) & -      & -      & -      & -      & - & - & 8  & 8 (120)   & 16  \\
				MF     & 9 (40) & 9 (9)   & 5 (19) & 20 (22)& -      & 1 (11) & - & - & 44 & 44 (101)   & 88  \\
				\midrule
				\textbf{Total} & \textbf{9 (40)} & \textbf{23 (169)} & \textbf{11 (93)} & \textbf{45 (64)} & \textbf{2 (18)} & \textbf{1 (11)} & \textbf{2} & \textbf{2} & \textbf{87} & \textbf{91 (395)} & \textbf{178} \\
				\bottomrule
			\end{tabular}
		}
			\begin{flushleft}
				\footnotesize
				\textbf{Note:} Numbers in () is LoM of elements extended with \profile; 
				attr: attribute; occur: occurrence; ISR: InsufficientResolution; MI: MissingInfo; ND: Non-determinism; IndSpec: IndeterminacySpecification.
			\end{flushleft}
		\end{minipage}%
	}
\end{table}

In addition, Table~\ref{tab:indeterminacy_source} presents identified sources of uncertainty, i.e., indeterminacy sources. 
In total, 91 elements were annotated with «IndeterminacySource», yielding 178 places where «IndeterminacySpecification» were applied, which characterize different nature of indeterminacy, including \textit{Insufficient Resolution}, \textit{Missing Information}, and \textit{Non-determinism}.
Indeterminacy modeling shows a similarly structured distribution as uncertainty.
MF (44 sources, 88 specifications) and DS (8 sources, 16 specifications) extensively apply indeterminacy at the structural levels to determine their operational status.
VM contributes 13 sources and 26 specifications at the structural level, specifically on \textit{port def} and \textit{attribute} elements, whereas AF and IS contribute smaller sets applied to communication-related elements, such as \textit{port def} and \textit{port}.
VFEA does not include explicit indeterminacy sources, as the uncertainty is associated with the \textit{wheelDiameter} attribute of \emph{Vehicle} (see Section~\ref{subsubsec:uncertainty}). 
This uncertainty arises from inherent and irreducible real-world variability, such as manufacturing tolerances, tire wear, and thermal expansion, which are beyond the scope of the analysis model and are therefore not further specified.
Regarding the nature of indeterminacy sources, non-determinism is the dominant, i.e., 87 out of the 91 sources across six case studies. 
In contrast, missing information and insufficient resolution occur in more specific contexts, with 2 instances of missing information in VM and 2 instances of insufficient resolution in ACC.
This may suggest that most modeled uncertainties in the selected systems are driven by variability in operational status and interaction reliability, whereas information incompleteness and resolution limits are applied in specific analysis or perception-related situations.

In summary, \profile is capable of characterizing uncertainty across all seven case studies, covering both structural and behavioral modeling elements. 
It supports the explicit representation of uncertainty in attributes, ports, transitions, actions, and messages, while also capturing their underlying indeterminacy sources and associated specifications. 
This may demonstrate the flexibility and applicability of our \profile across diverse modeling contexts, ranging from analytical models to operational and interaction-driven models. 
Overall, the results confirm that \profile provides a systematic way for identifying, characterizing, and analyzing uncertainty within \sysml models.

\subsection{Discussion}\label{subsec:discussion}

\textbf{Enable uncertainty modeling in \sysml.}
Applying \profile not only annotates existing model elements, but also introduces additional modeling constructs to explicitly and systematically represent \textit{uncertainty} and its associated characteristics.
These constructs include belief statement, indeterminacy sources, indeterminacy specifications, and measurement definitions, which together provide a structured means to capture the scope, origin, nature, and quantification of uncertainty. 

\begin{figure}
\begin{lstlisting}[escapeinside={(*@}{@*)}]
«Uncertainty<ocr, epi, subj>» transition engageDefense
	first InGrid
	accept defenseCommand : Domain::ShipCommand via miningFrigates.controlPort
	do action engageDefenses : EngageDefenses
	then InGrid {
		u_reducibility = PartiallyReducible;
		u_pattern = Random;
		
		«IndeterminacySpecification» ref constraint miningFrigateControlportOperational ::> miningFrigates.controlPort.Operational;
		«IndeterminacySpecification» ref constraint shieldModuleOperational ::> miningFrigates.shieldModule.Operational;
		«IndeterminacySpecification» ref constraint droneBayOperational ::> miningFrigates.droneBay.Operational;
		«IndeterminacySpecification» ref constraint defenseTurretOperational ::> miningFrigates.defenseTurret.Operational;
		«IndeterminacySpecification» constraint overallSpecification {
			miningFrigateControlportOperational and shieldModuleOperational and droneBayOperational and defenseTurretOperational
		}
	}
(*@\textellipsis@*)
«Uncertainty<ocr, epi, subj>» transition failToEngageDefense
	first InGrid
	accept defenseCommand : Domain::ShipCommand via miningFrigates.controlPort
	then InGrid {
		u_reducibility = PartiallyReducible;
		u_pattern = Random;
		
		«IndeterminacySpecification» ref constraint miningFrigateControlportNotOperational ::> miningFrigates.controlPort.NotOperational;
		«IndeterminacySpecification» ref constraint shieldModuleNotOperational ::> miningFrigates.shieldModule.NotOperational;
		«IndeterminacySpecification» ref constraint droneBayNotOperational ::> miningFrigates.droneBay.NotOperational;
		«IndeterminacySpecification» ref constraint defenseTurretNotOperational ::> miningFrigates.defenseTurret.NotOperational;
		«IndeterminacySpecification» constraint overallSpecification {
			miningFrigateControlportNotOperational or shieldModuleNotOperational or droneBayNotOperational or defenseTurretNotOperational
		}
		
		metadata defenseEngagementFailureRisk : RiskMetadata::Risk {
			totalRisk {
				impact = RiskMetadata::LevelEnum::high;
			}
		}
	}
\end{lstlisting}
\caption{Model snippet illustrating additional uncertain transitions in MF}
\label{lst:mf_engageDefense}
\end{figure}

\begin{figure}
\begin{lstlisting}[escapeinside={(*@}{@*)}]
«IndeterminacySource<nd>» occurrence def NonDeterministicComponent {
	attribute operationalStatus : Boolean;
	
	«IndeterminacySpecification» constraint Operational {
		operationalStatus;
	}
	
	«IndeterminacySpecification» constraint NotOperational {
		not operationalStatus;
	}
}

port def PodPort specializes NonDeterministicComponent { // inherent IndeterminacySource and IndeterminacySpecifications
	in item command : ShipCommand;
	out item report : ShipReport;
}
(*@\textellipsis@*)
part def DroneBay specializes NonDeterministicComponent { // inherent IndeterminacySource and IndeterminacySpecifications
	attribute maxDrones : Integer; // Maximum drones stored
}
\end{lstlisting}
	\caption{Model snippet illustrating indeterminacy sources in MF}
	\label{lst:mf_nd_sources}
\end{figure}

For example, as shown in Figure~\ref{lst:mf_engageDefense}, MF defines two alternative uncertain transitions (\textit{engageDefense} and \textit{failToEngageDefense}), which are triggered by the same incoming command: \textit{defenseCommand}. 
Transition usage \textit{engageDefense} represents the successful defense engagement path, while transition usage \textit{failToEngageDefense} captures the failure path under adverse operational conditions that we further introduced when applying \profile. 
In the successful transition, the overall condition is defined as a conjunction of operational constraints, requiring all relevant components, such as the control port, shield module, drone bay, and defense turret, to be operational. 
In contrast, the failed transition is defined as a disjunction of non-operational constraints, where the failure of any one of these components is sufficient to prevent successful defense engagement. 
This explicit indeterminacy specification makes the causal assumptions underlying the system behavior visible and traceable within the model.

The corresponding indeterminacy sources and their operational and non-operational specifications are shown in Figure~\ref{lst:mf_nd_sources}. 
We define ocurrence definition \textit{NonDeterministicComponent}, stereotyped with «IndeterminacySource<nd>», to capture the inherent non-determinism of component operational status through attribute \textit{operationalStatus}.  
This definition can be specialized by structural elements such as \textit{PodPort} and \textit{DroneBay}, allowing their operational conditions to be explicitly referenced by «IndeterminacySpecification» constraints in behavioral transitions. 
This establishes a clear and traceable link between structural indeterminacy sources and behavioral uncertainty manifestations.

The uncertainty profile also enables the explicit characterization of uncertainty properties. 
As shown in Figure~\ref{lst:mf_engageDefense}, both transitions are annotated with {«Uncertainty<ocr, epi, subj>»} and further characterized using attributes such as \textit{u\_reducibility} and \textit{u\_pattern}. 
This allows uncertainty to be precisely classified in terms of its reducibility and probabilistic nature, rather than being treated as an informal or implicit assumption.
Furthermore, the profile enables explicit modeling of uncertainty consequences.
The failed transition is annotated with the risk metadata element \textit{defenseEngagementFailureRisk}, which specifies a high impact level (see Figure~\ref{lst:mf_engageDefense}). 
This establishes explicit traceability between indeterminacy sources, uncertain behaviors, and their potential system-level impacts, allowing the consequences of uncertainty to be represented and analyzed directly within the model. 

Overall, \profile improves the expressiveness of uncertainty modeling by enabling a comprehensive, explicit, and analyzable representation of uncertainty, its underlying causes, and its potential effects.

\textbf{Enable uncertainty propagation analysis.}
With \profile, it enables explicit modeling of how uncertainty propagates across interacting components, actions, and behavioral transitions. 
By linking uncertain behaviors, indeterminacy specifications, and their downstream effects, the profile establishes a traceable propagation chain that captures how upstream uncertainty can influence subsequent system behavior and potentially lead to failure conditions.

\begin{figure}
\begin{lstlisting}[escapeinside={(*@}{@*)}]
«Uncertainty<ocr, epi, subj>» transition failToAcceptCallGiveItems
	first WaitOnData
	accept cl:CallGiveItems via tellu.APIS_HTTP
	then WaitOnData {
		u_reducibility = PartiallyReducible;
		u_pattern = Random;
		
		«IndeterminacySpecification» ref constraint portNotOperational ::> tellu.APIS_HTTP.APIS_HTTP_Not_Operational;
		«IndeterminacySpecification» ref constraint consumerPortNotOperational ::> TellUConsumer.apisp.APIS_HTTP.APIS_HTTP_Not_Operational;
		«IndeterminacySpecification» constraint overallSpecification {
			portNotOperational or consumerPortNotOperational
		}
		
		«Effect» ref ::> TellUConsumer.TellUbehavior.failToAcceptResultGiveItems;
		
		metadata lossOfCallGiveItemsRisk : RiskMetadata::Risk {
			totalRisk {
				impact = RiskMetadata::LevelEnum::medium;
			}
		}
	}
(*@\textellipsis@*)
«Uncertainty<ocr, epi, subj>, Effect» transition failToAcceptResultGiveItems
	first Wait
	accept rs:ResultGiveItems
	then Wait {
		u_reducibility = PartiallyReducible;
		u_pattern = Random;
		
		«IndeterminacySpecification» ref constraint portNotOperational ::> apisp.APIS_HTTP.APIS_HTTP_Not_Operational;
		«IndeterminacySpecification» ref constraint producerPortNotOperational ::> APISProducer.tellu.APIS_HTTP.APIS_HTTP_Not_Operational;
		«IndeterminacySpecification» constraint overallSpecification {
			portNotOperational or producerPortNotOperational
		}
		
		metadata resultReceptionFailureRisk : RiskMetadata::Risk {
			totalRisk {
				impact = RiskMetadata::LevelEnum::high;
			}
		}
	}
\end{lstlisting}
	\caption{Model snippet illustrating uncertainty-to-uncertainty propagation in AF}
	\label{lst:af_fail_publish}
\end{figure}

As shown in Figure~\ref{lst:af_fail_publish}, transition \textit{failToAcceptCallGiveItems} is annotated with «Uncertainty<ocr, epi, subj>», representing uncertain behavior of the system to accept an incoming \textit{CallGiveItems} request via the \textit{APIS\_HTTP} communication port. 
This uncertainty is explicitly linked to indeterminacy specifications, i.e., \textit{portNotOperational} and \textit{consumerPortNotOperational}, which capture the non-operational status of the communication interface on either the receiving or consuming component. 
These non-deterministic conditions define the potential causal factors that may prevent the successful handling of the incoming call. 
As a result of this failure, the intended downstream action to send the corresponding \textit{ResultGiveItems} response is not performed.
Instead, the model explicitly references the downstream behavioral effect \textit{TellUConsumer.TellUbehavior.failToAcceptResultGiveItems}, indicating that upstream uncertainty may propagate to subsequent system behavior.

This propagated uncertainty is further reflected in transition \textit{failToAcceptResultGiveItems}, which is annotated with both «Uncertainty» and «Effect». 
This transition captures the failure of the system to receive the \textit{ResultGiveItems} message due to non-operational communication ports due to non-operational communication interfaces on either the producer or consumer side, as linked to \textit{portNotOperational} and \textit{producerPortNotOperational}. 
This causal-effect linkage establishes a traceable causal propagation chain, where upstream uncertainty in accepting the call propagates to downstream failure in handling the result.
Furthermore, the inclusion of risk metadata, such as \textit{lossOfCallGiveItemsRisk} and \textit{resultReceptionFailureRisk}, further characterizes the system-level consequences of these propagated uncertainties by specifying their associated impact levels.

Such explicit modeling structure supports both forward and backward uncertainty propagation analysis. 
Forward analysis enables tracing how an upstream uncertain cause, such as a non-operational communication port, can propagate through interactions and lead to downstream uncertain behaviors or failure transitions. 
Conversely, backward analysis enables tracing from an observed failure, such as \textit{failToAcceptResultGiveItems}, to its originating indeterminacy sources and uncertain upstream actions. 
By making these propagation paths explicit within the model, \profile improves the traceability, diagnosability, and analyzability of uncertainty propagation across interacting systems, to enhance uncertainty-aware system analysis and failure reasoning.

\textbf{Enable uncertainty analyses via uncertainty topics.}
\profile enables flexible uncertainty analysis through «UncertaintyTopic», which allows multiple uncertainty-related model elements to be grouped under a shared analytical concern. 
Rather than analyzing each uncertainty in isolation, «UncertaintyTopic» provides a structured mean to organize uncertainties according to functional, operational, or architectural contexts.

As shown in Figure~\ref{lst:af_uncertainty_topics}, uncertainty topics such as \textit{PublishTopic} and \textit{RemoteCallTopic} group uncertain actions and transitions associated with specific operational concerns. 
For example, \textit{PublishTopic} links the publishing action \textit{sendPublish}, the normal reception behavior \textit{acceptPublish}, and the failure transition \textit{failToAcceptPublish}, explicitly capturing the uncertainty associated with message publishing. 
This grouping makes explicit the relationships between uncertainty sources, behavioral manifestations, and failure outcomes.

\begin{figure}
\begin{lstlisting}[escapeinside={(*@}{@*)}]
«UncertaintyTopic» item def PublishTopic {
    «Uncertainty» ref ::> AHFNorway_LocalCloudDD.APISProducer.APISPbehavior.sendPublish;
    «Uncertainty» ref ::> AHFNorway_LocalCloudDD.MQTTServer.Serve.acceptPublish;
    «Uncertainty» ref ::> AHFNorway_LocalCloudDD.MQTTServer.Serve.failToAcceptPublish;
}
(*@\textellipsis@*)
«UncertaintyTopic» item def RemoteCallTopic {
	«Uncertainty» ref ::> AHFNorway_LocalCloudDD.TellUConsumer.TellUbehavior.sendCallGiveItems;
	«Uncertainty» ref ::> AHFNorway_LocalCloudDD.APISProducer.APISPbehavior.acceptCallGiveItems;
	«Uncertainty» ref ::> AHFNorway_LocalCloudDD.APISProducer.APISPbehavior.failToAcceptCallGiveItems;
	«Uncertainty» ref ::> AHFNorway_LocalCloudDD.TellUConsumer.TellUbehavior.acceptResultGiveItems;
	«Uncertainty» ref ::> AHFNorway_LocalCloudDD.TellUConsumer.TellUbehavior.failToAcceptResultGiveItems;
}
\end{lstlisting}
\caption{Model snippet illustrating uncertainty-topic grouping in AF}
\label{lst:af_uncertainty_topics}
\end{figure}

This topic-based organization supports flexible analysis at multiple levels. 
At the element level, modelers can perform fine-grained diagnosis by tracing causes and effects of individual uncertain actions, transitions, or messages. 
At the topic level, modelers can assess the overall impact of uncertainty on specific operational concerns, enabling systematic impact assessment. 
Topic grouping also enables cross-topic comparisons to identify which functional areas are most affected by uncertainty. 
In addition, «UncertaintyTopic» supports uncertainty analysis across different stages of the system development lifecycle, including requirements, design, implementation, and testing. 
For example, uncertainties identified in requirements or design models can be traced to their behavioral manifestations in operational models and further linked to verification or testing concerns, enabling end-to-end traceability and consistency of uncertainty analysis throughout system development.

By organizing uncertainty-related elements into coherent analytical groups, \profile improves the ability to perform various analysis, such as, impact analysis, failure diagnosis, consistency analysis, and uncertainty-driven reasoning across interacting system components. 

\textbf{Require tool support.}
Using this profile effectively requires dedicated tool support. 
In the current setting, most modeling work is performed in textual form, which limits immediate usability for large-scale industrial modeling tasks and increases manual effort.
In particular, important assistance capabilities are still limited, including automated constraint validation and consistency checking across uncertainty annotations, indeterminacy sources/specifications, effects, and related model elements. 
As a result, model quality currently depends heavily on manual review, and intelligent tool-assisted checks are needed to improve reliability, scalability, and adoption.
To address these limitations, we are developing dedicated tool support for \profile, built on ModelCopilot\textsuperscript{\ref{foot:mc}} platform,
a novel model-based engineering platform that already provides an implementation of \sysml. 

\subsection{Threats to validity}\label{subsec:threats}

\textbf{Internal validity.}
Model extensions were manually introduced and reviewed, which may introduce modeling bias, such as variations in uncertainty identification and annotation across case studies. 
To mitigate this risk and support transparency and reproducibility, we have open-sourced both the original models and the extended models, enabling independent review, replication, and further evaluation.

\textbf{External validity.}
Due to limitations of availability of \sysml models, this study evaluates \profile using seven case studies obtained from two public repositories. 
Although these case studies cover diverse domains and vary in model size and complexity, they may not fully represent all modeling practices or domain-specific contexts. 
As a result, the generalizability of the findings may be limited. 
Further empirical study using additional models, when available, is needed to strengthen confidence in the applicability of our \profile across broader modeling contexts.

\section{Related Work}\label{sec:related_work}

Numerous studies have proposed solutions for modeling/specifying uncertainties by employing a range of notations and formalisms. Below, we discuss some of them.

Probabilistic models have been used to specify uncertainty.
For instance, Geng et al.~\cite{geng2017modeling} used possibility spatio-temporal hybrid automata models to describe CPS behaviors.
Agli et al.~\cite{agli2016business} used probabilistic relational models to represent uncertain values to enable probabilistic reasoning in business rule management systems.
Burgueño et al.~\cite{burgueno2019specifying} used probability distributions and defined an extension of basic type Real to provide an algebra of operations for specifying measurement uncertainty.
When probability distributions cannot be established due to the lack or poor quality of samples, intervals can be used. For instance, Vallecillo et al.~\cite{vallecillo2016expressing} proposed to extend UML and OCL types to specify data uncertainty, which supports the use of intervals to represent the ranges of possible values.
%
Fuzzy set theory~\cite{zimmermann2011fuzzy} is also commonly used. For instance, Sicilia et al.~\cite{sicilia2004extending} proposed to perform fuzzy modeling with UML. Han et al.~\cite{han2016fame} proposed a UML-based solution for modeling fuzzy self-adaptive software. Ma et al.~\cite{ma2012modeling} investigated fuzzy data modeling on UML and XML data models.


Furthermore, Sedaghatbaf and Abdollahi Azgomi~\cite{sedaghatbaf2018reliability} extended UML to specify uncertain parameter in UML models.
Burgueno et al.~\cite{burgueno2019belief} proposed a belief UML profile, which was later extended in~\cite{burgueno2023dealing} to represent agents' beliefs their associated uncertainty in UML models, based on subjective logic~\cite{munoz2020extending}.
Zhang et al.~\cite{zhang2019uncertainty} proposed the UML Uncertainty Profile, based on the uncertainty conceptual model U-Model~\cite{zhang2016understanding}, as part of the \textit{UncerTum} test modeling framework to enable uncertainty-aware testing of CPS, i.e., UncerTest~\cite{zhang2019UncerTest}.
Zhang et al. proposed to specify uncertainty requirements on use case models~\cite{zhang2018specifying}.

SysML gained less attention in uncertainty modeling as compared with UML.
Jongeling et al.~\cite{jongeling2023uncertainty} proposed annotating degrees of doubt (belief, disbelief, and uncertainty) in subjective logic to indicate design uncertainties on SysML v1 model elements.
Cvijic et al.~\cite{cvijic2025probabilistic} proposed ProbSysML, a plugin for Cameo, to enable the annotation of probabilities to SysML v1's component attributes, and defined probabilistic relationships between components and their corresponding parts.

These proposals are often tailored to specific applications, and without a shared conceptual base, which inevitably leads to issues such as inconsistent terminology, conflicting notations, and duplicated efforts~\cite{troya2021uncertainty}.
To this end, Yue et al. initiated the definition of the Precise Semantics for Uncertainty Modeling (PSUM)~\cite{omg2024psum} standard in OMG, based on U-Model~\cite{zhang2016understanding}. PSUM provides an unified metamodel for representing, characterizing and measuring belief, indeterminacy and uncertainty, as the first international standard on uncertainty modeling.
While PSUM provides a solid metamodel foundation, it is intentionally designed to be independent of any specific modeling language. Consequently, its integration and realization within particular modeling domains remain to be systematically established.
Hence, in this work, we propose \profile to systematically extend \sysml to provide a standardized alignment for representing uncertainty in system modeling.

\section{Conclusion and Future Work}\label{sec:conclusion}

Handling uncertainty poses a significant challenge due to the inherent complexity of systems such as microservice-based systems and autonomous systems.
The first and critical step is to model uncertainty explicitly and systematically in any Model-Based Systems Engineering (MBSE) practice.

In this paper, we integrated the state-of-the-art standards PSUM and \sysml to establish a systematic modeling approach for explicitly characterizing uncertainty in next-generation system modeling practices.
Our approach combines the strengths of PSUM and \sysml to enhance the expressiveness of uncertainty modeling in a precise and structured manner, aiming to enable uncertainty-aware MBSE, uncertainty analysis, uncertainty propogation, etc.
To evaluate \profile, we conducted seven case studies collected from the literature.
Results confirm the feasibility and applicability of the proposed profile for representing uncertainties spanning diverse modeling contexts, ranging from structural models to event-driven behavioral models.
Through these case studies, we observed that applying \profile potentially helps to support the identification of additional uncertainties, improve the transparency and traceability of uncertainty sources and effects, and enable further advanced uncertainty-aware analyses and workflows, such as uncertainty propagation and risk analysis.

For future work, we plan to integrate \profile within the ModelCopilot platform to provide enhanced tool support and strengthen the modeling community.
We plan to develop automated validation and checking approaches to ensure the correctness, consistency, and completeness of uncertainty models.
In addition, we will investigate automated suggestion and model generation strategies using Language Models (LMs) to more effectively assist engineers in identifying, construing and evolving uncertainties.
Furthermore, we will explore novel uncertainty analysis approaches that support multiple uncertainty viewpoints, allowing stakeholders to analyze and assess uncertainty from structural, behavioral, and operational perspectives.
Last but not least, we plan to develop management techniques for uncertainty models and their holders (i.e., belief agents) to enhance model traceability over time, and to support change impact analysis and consistency maintenance as systems and their associated uncertainties evolve.

\section*{Acknowledgements}

This work is supported by the National Natural Science Foundation of China (grant agreement No. 62502022).
%
\section*{Declarations}


\begin{itemize}
 \item \textbf{Funding}. This work is funded by the National Natural Science Foundation of China (grant agreement No. 62502022).
 \item \textbf{Author contribution.}
 All authors contributed to conceptualization, and the design of the profile and case studies.
 Man Zhang conducted the analysis of the case studies, and prepared the initial drafts of the case studies, conclusion and future work sections.
 Yunyang Li implemented the profile, carried out the case studies, collected the corresponding results, and prepared the initial draft of the related work and methodology sections.
 Tao Yue drafted the first versions of the abstract, introduction, and background sections.
 All authors participated in revising and refining the manuscript, and approved the final version.
 \item \textbf{Conflict of interest.} The authors declared that they have no conflict of interest.
 \item \textbf{Data availability.} \profile profile and raw data of the case studies (i.e., original \sysml models and models applied with \profile) are available online at \url{https://github.com/WSE-Lab/PSUM-SysMLv2}.
 \item \textbf{Ethics approval.} Not applicable.
 \item \textbf{Informed consent.} Not applicable.
 \item \textbf{Clinical trial number.} Not applicable.
\end{itemize}

\bibliographystyle{ACM-Reference-Format} 

%

\bibliography{sn-bibliography}

\newpage

\end{document}